\documentclass[12pt,a4paper,twoside]{article}
\usepackage{epsf,epsfig,amsfonts,amsmath,amssymb,axodraw,feynmf}

\date{}

\begin{document}

\centerline{}

\centerline{}

\centerline {\Large{\bf Tetrahedron Diagram and Perturbative
Calculation}}

\centerline{}

\centerline{\Large{\bf in Chern-Simons-Witten Theory}}

\centerline{}

\centerline{\bf {Freddy P
Zen$^{a,b,}$\footnote{fpzen@fi.itb.ac.id}, Jusak S
Kosasih$^{a,b,}$\footnote{jusak@fi.itb.ac.id}, Asep Y
Wardaya$^{a,c,}$\footnote{asepyoyo@yahoo.co.id},
Triyanta$^{a,b,}$\footnote{triyanta@fi.itb.ac.id}}}

\centerline{}

\centerline{$^a$Theoretical Physics Laboratory, THEPI Division, }

\centerline{and}

\centerline{$^b$INDONESIA Center for Theoretical and Mathematical
Physics (ICTMP),}

\centerline{Faculty of Mathematics and Natural Sciences, Institut
Teknologi Bandung,}

\centerline{Jl. Ganesha 10 Bandung 40132, Indonesia.}
\centerline{}

\centerline{$^c$Diponegoro University,}

\centerline{ Jl. Prof. H. Soedarto, SH. Tembalang Semarang 50275,
Indonesia.}

\newtheorem{Theorem}{\quad Theorem}[section]

\newtheorem{Definition}[Theorem]{\quad Definition}

\newtheorem{Corollary}[Theorem]{\quad Corollary}

\newtheorem{Lemma}[Theorem]{\quad Lemma}

\newtheorem{Example}[Theorem]{\quad Example}

\begin{abstract}
We investigate extended Wilson loop operators, in particular
tetrahedron operator in $(2+1)$-dimensional Chern-Simons-Witten
theory. This operator emerges naturally from the contribution
terms in two-particle scattering amplitude. We evaluate this
diagram non-perturbatively in terms of vacuum expectation values
of Wilson loop operators, especially for gauge group $\mathop{\rm
SU}(N)$ with specific choices of representations. On the other
hand, we also discuss the perturbative calculation of vacuum
expectation value in this theory. We show that, up to the third
order, this values of unknotted Wilson loop operators are
identical to the non-perturbative result.

\end{abstract}


\section{Introduction}
The Chern-Simons-Witten (CSW) theory has been providing many
interesting topics for both mathematics and physics. Different
aspects of the theory have been explored, e.g. abelian and
non-abelian theory with the Chern-Simons term, supersymmetric
extension, and pure theory. Specifically, after it was pointed out
that the pure theory could be relevant for knot theory
\cite{Schwarz1978,Atiyah1988}, many researchers have contributed
to clarify the relation of CSW theory to two-dimensional,
conformal field theories and to knot theory. Witten described the
exact solution to this theory in non-perturbative case. He and
others also showed that there is relation between this theory and
the polynomial or quantum group invariants of knot in three
dimensions, in particular the Jones polynomial and its
generalizations like the HOMFLY and the Kauffman polynomial
invariants \cite{Jones1985,FreydYetterHoste1985,Witten1989a}.
These knot polynomials can be regarded as vacuum expectation
values (VEV) of Wilson loop operators in this theory.

The CSW theory also has interesting features from the perturbative
point of view and lead to Goussarov-Vassiliev or finite type
invariants. Moreover, Witten's idea was based on the validity of
the path integral formulation of the quantized theory. By
examining the existence of a quantum field theory description of
the link invariants in perturbative framework, it turned out that
the coefficients of the perturbative series correspond to these
invariants. The perturbative expansion of the path integral of the
theory has been considered for flat $R^3$
\cite{GuadagniniMartelliniMintchev1989} and for general
three-manifolds \cite{AxelrodSinger1991}. Subsequently, there was
much attention given to understanding the perturbative series for
knots and links in $R^3$ \cite{AltschulerFreidel1997}.

The perturbative series expansion has been studied for different
gauge-fixings which lead to different representations for
Vassiliev invariants. The covariant Landau gauge corresponds to
the configuration space integrals and the non-covariant light-cone
gauge to the Kontsevich integrals. Another studies of the
perturbative series expansion in the non-covariant temporal gauge
has the important feature that the integrals which are present in
the expressions for the coefficients of the perturbative series
expansion can be carried out. In this case one obtains
combinatorial expressions, instead of integral ones, for Vassiliev
invariants and this has been shown to be the case up to order four
\cite{LabasPerez}. Perturbative expansion in other gauges might
also highlight other aspects of the theory, for example in axial
gauge \cite{Hahn2004} and in light-cone gauge
\cite{TomoshiroOchiai}.  The invariants obtained in the
perturbative framework with different gauge-fixing are the same
since the theory is gauge invariant and Wilson loops are gauge
invariant operators.

One main issue in perturbative CSW theory is the calculability of
the vacuum expectation value of Wilson loop operator $\langle
W(C)\rangle$ in the three-dimensional field theory framework. It
turned out that $\langle W(C)\rangle$ has a meaningful
perturbative expansion in powers of the coupling constant
$k^{-1}$. Although by power counting the theory appears
renormalizable, it is in fact UV finite, which means the $\beta$
function and the anomalous dimensions of the fields vanish to all
orders \cite{BlasiCollina1990,DelducLucchesiPiguetSorella1990}.
Even, more interestingly, there is no divergences in the
computation of $\langle W(C)\rangle$ in this theory. One loop
renormalization constant of the theory has been calculated and the
result showed the existence of the famous $k$ shift
\cite{ChenZhu1993}. Therefore, the framing has nothing to do with
divergences of $\langle W(C)\rangle$, but is related to the
self-linking problem which is topological in origin.

The mathematical establishment of relations between CSW gauge
theory, topological theories in 3 manifold and knot theory is
still making progress. Some recent studies include the application
of Maldacena's conjecture \cite{HirosiVafa}, the use of methods of
stochastic analysis \cite{AlbeverioMitoma2007} and its connection
to the Penner models \cite{Chair2007}.

In this paper we will evaluate the non-perturbative as well as the
perturbative aspects of CSW theory. In non-perturbative aspect, it
is shown that the VEV of Wilson loop operators are evaluated by
using braiding formula \cite{Hayashi1993} which is useful to
construct algebraic relations between unknotted Wilson loop
operators. It also discussed the emergence of the extended Wilson
loop operators, namely baryon type \cite{Witten1989b} and
tetrahedron operators \cite{Zen1994}. Especially, we will evaluate
the tetrahedron type operator. This diagram emerges by refining
the calculation of the gravitational scattering amplitude in
previous work \cite{KoehlerMansouriVazWitten1991}. It is
important, therefore, to make it clear what contributions will be
supplied from these terms.

On the other hand, parallel to the non-perturbative approach, the
explicit perturbative calculation of the unknotted Wilson loop
operator is presented up to the third order. The coefficients of
these expansion is shown to be the same as the non-perturbative
results. After that, we also investigate the ghost and auxiliary
fields contributions to the Wilson loop operator. A detailed
perturbative calculation of the ghost contribution up to the
second order is given in the Appendix A and other important
integral formulas related to the framing procedure are given in
Appendix B.

\section{Chern-Simons Theory and Extended Wilson Loop Operators}
In this section we will present the rudimentary facts about CSW
theory and to discuss the emergence of extended Wilson loop
operators\footnote{One of us (FPZ) would like to thank M. Hayashi
for collaboration in this part, see ref. \cite{Zen1994}}. The
action for this theory in $(2+1)$-dimension is the Chern-Simons
secondary characteristic class defined by
\begin{equation}
\label{scsw}
    S=\frac{k}{4\pi} \int_M  \mbox{Tr} (A\wedge dA + \frac{2}{3} A\wedge A\wedge A)  .
\end{equation}
Here $k^{-1}$ plays the role of coupling constant whereas $A$ is a
connection on a $G$-bundle $E$ over a space-time three manifold
$M$. Trace is taken over the representation of gauge group $G$.
The partition function $Z(M,k)$ takes the form
\begin{equation}
    Z(M,k)= \int_M  [DA] \; e^{iS} .
\end{equation}
Under the gauge transformation, the action will transform as
\begin{equation}
    S \longrightarrow S + 2\pi k \; S_{WZ} ,
\end{equation}
where $S_{WZ} \in \mathbb{Z}$  is the winding number of the gauge
transformation.

The Wilson loop operators $W_{\rho}(C)$, which are the basic
observables of the theory, are the most important gauge invariant
operators in this theory. These operators are related to the link
invariants of knot theory and defined by the trace of path-ordered
exponential of the connection $A$ along a closed loop $C$ which is
embedded in $M$,
\begin{equation}
    W_{\rho}(C) = \mbox{Tr}_{\rho} \left(\mathcal{P} \exp \oint_C A\right) ,
\end{equation}
where $\rho$ is a representation of the gauge group $G$
($\bar{\rho}$ is its conjugate representation). Note that a closed
loop $C$ can be knotted in the three manifold.

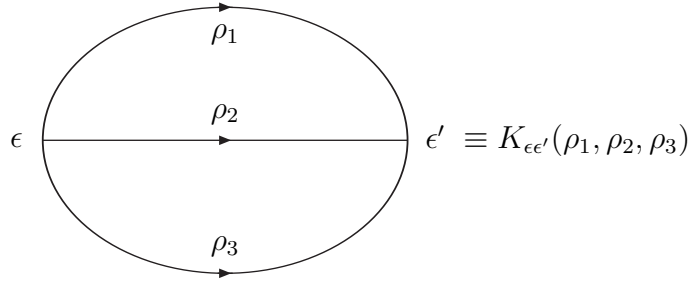
\begin{figure}[t]
\parbox{15cm}{
\begin{picture}(200,100)(-150,-50)

\ArrowLine(-68,0)(68,0) \Text(-78,0)[c]{$\epsilon$}
\Text(75,0)[l]{$\epsilon' \; \equiv
K_{\epsilon\epsilon'}(\rho_1,\rho_2,\rho_3)$}
\Text(0,40)[c]{$\rho_1$} \Text(0,10)[c]{$\rho_2$}
\Text(0,-40)[c]{$\rho_3$}

\Oval(0,0)(50,68)(0) \ArrowArcn(0,-30)(80,92,88)
\ArrowArc(0,30)(80,268,272)

\end{picture}
} \caption{A baryon-type operator} \label{fig:baryon}
\end{figure}

A normalized VEV of an operator $O(A)$ is defined as
\begin{equation}
    \langle O(A) \rangle = \frac{Z(M,k,O)}{Z(M,k,1)} ,\hspace{1cm}  Z(M,k,O)=\int [DA] \; O(A) e^{iS} .
\end{equation}
In the following we use notation for the normalized VEV of
unknotted Wilson loop operator

\begin{equation}
    E_0(\rho)=\langle W_{\rho}(\circlearrowleft)\rangle ,
\end{equation}
such that the VEV of the baryon-type operator
(Figure~\ref{fig:baryon}) is denoted as
\begin{equation}
    \langle K_{\epsilon\bar{\epsilon}'}(\rho_1,\rho_2,\rho_3)\rangle=\delta_{\epsilon\bar{\epsilon}'}\sqrt{E_0(\rho_1)E_0(\rho_2)E_0(\rho_3)} ,
\label{vev-baryon}
\end{equation}
and if, for example $\rho_1={\rm id}$ (identity representation),
then from consistency condition \cite{Witten1989b}
\begin{equation}
    \langle K_{\epsilon\bar{\epsilon}'}({\rm id},\bar{\rho_3},\rho_3)\rangle=\delta_{\epsilon\bar{\epsilon}'}E_0(\rho_3) .
\end{equation}

\begin{figure}[t]
\parbox{15cm}{
\begin{picture}(200,200)(-200,-100)

\Line(-60,100)(100,100) \Line(-100,60)(60,60)
\Line(-100,60)(-60,100) \Line(100,100)(60,60)

\Line(-60,-60)(100,-60) \Line(-100,-100)(60,-100)
\Line(-100,-100)(-60,-60) \Line(100,-60)(60,-100)

\DashLine(0,80)(0,60)2 \Line(0,60)(0,42) \Line(0,40)(0,17)
\Line(0,14)(0,5)


\Line(0,-20)(0,-30) \Line(0,-32)(0,-55) \ArrowLine(0,-85)(0,-58)
\ArrowLine(-15,-70)(-12,-70)

\qbezier[10](-30,80)(-30,70)(-20,60)
\qbezier(-20,60)(-10,55)(-2,54) \qbezier(2,54)(40,49)(-5,40)
\qbezier(-5,40)(-30,35)(-2,30) \qbezier(2,29)(40,24)(-5,15)
\qbezier(-5,15)(-30,10)(-2,5)

\qbezier[5](-5,0)(-5,-7)(-5,-15)

\qbezier(-30,-80)(-30,-70)(-3,-70) \qbezier(3,-70)(30,-68)(-5,-55)
\qbezier(-5,-55)(-40,-46)(-2,-45) \qbezier(2,-45)(35,-39)(-5,-30)
\qbezier(-5,-30)(-40,-23)(-2,-20)

\LongArrow(80,0)(110,0) \LongArrow(80,0)(80,30)
\Text(115,0)[c]{$\Sigma$} \Text(85,30)[c]{$t$}

\Text(0,80)[c]{$\times$} \Text(0,-85)[c]{$\times$}
\Text(-45,-85)[c]{$|i\rangle$} \Text(-45,85)[c]{$\langle f|$}
\Text(80,92)[c]{$\Sigma_f$} \Text(80,-68)[c]{$\Sigma_i$}
\Text(10,-75)[c]{$\sigma$} \Text(-30,-70)[c]{$\rho$}
\Text(20,-5)[l]{$\left.\begin{array}{c} \\ \\ \\ \\ \\ \\ \\ \\
\end{array}\right\}g_{\rho\sigma}^{(2l)}$}
\Text(-30,-80)[c]{$\bullet$} \Text(-30,80)[c]{$\bullet$}
\end{picture}
} \caption{A test particle scattered off a massive source
particle} \label{fig:scattering}

\end{figure}
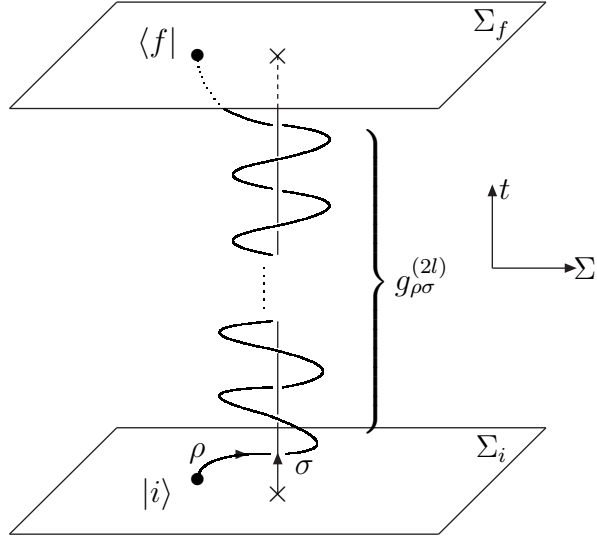

Now, let us imagine the following situation (see
Figure~\ref{fig:scattering}). A test particle scattered off a
massive source particle situated at the spatial origin of the
space-time manifold $M$. Upon quantization of the test particle,
the scattering amplitude describing this process becomes a sum of
contribution of homotopically inequivalent paths for the test
particle owing to the fact that the CSW theory is topological.

Such paths can be labelled by winding numbers between the
trajectories of the source and the test particles. If we project
the trajectories onto a surface $\Sigma$, then the process of
winding number $l$ is described by the braiding operator
$g_{\rho\sigma}^{(2l)}$. Here we assume that the representations
assigned to the source and test particles are $\rho$ and $\sigma$,
respectively. Consequently, the scattering amplitude is given by
Ref. \cite{Zen1994}
\begin{equation}
    \langle f | \sum^{\infty}_{l=-\infty}g_{\rho\sigma}^{(2l)} | i \rangle
    = \sum^{\infty}_{l=-\infty} \left\{ \frac{\langle \hat{g}_{\rho\sigma}^{(2l)} \rangle}{E_0(\rho)E_0(\sigma)}
      \langle f | g_{\rho\sigma}^{(0)} | i \rangle
    + \sum^{r-1}_{u=1} b_u^{(l)} \langle f | H_{\rho\sigma}(\phi_u) | i \rangle \right\}.
\label{scattering}
\end{equation}
The initial state $| i \rangle$ and the final state $\langle f |$
are connected by the operator $g_{\rho\sigma}^{(2l)}$, which
represents trajectories of the source and test particles.

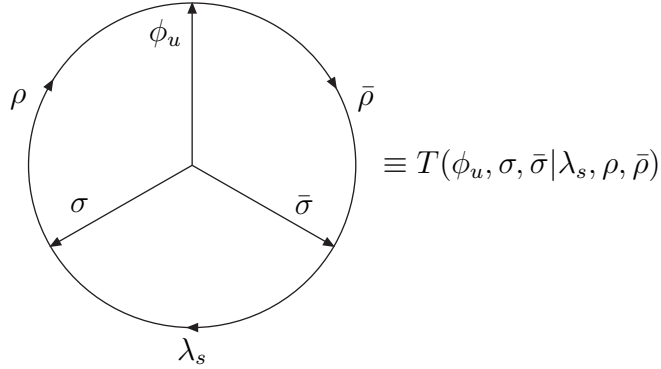
\begin{figure}[t]
\parbox{10cm}{
\begin{picture}(200,150)(-150,-80)
\ArrowArcn(0,0)(61,210,90) \ArrowArcn(0,0)(61,90,330)
\ArrowArcn(0,0)(61,330,210) \LongArrow(0,0)(0,60)
\LongArrow(0,0)(-52,-30) \LongArrow(0,0)(52,-30)
\Text(-10,50)[c]{$\phi_u$} \Text(-42,-15)[c]{$\sigma$}
\Text(42,-15)[c]{$\bar\sigma$} \Text(65,25)[c]{$\bar\rho$}
\Text(68,0)[l]{$ \; \equiv T(\phi_u,\sigma,\bar\sigma |
\lambda_s,\rho,\bar\rho)$} \Text(-65,25)[c]{$\rho$}
\Text(0,-70)[c]{$\lambda_s$}
\end{picture}
} \caption{A tetrahedron operator which contribute to scattering
amplitude of a test particle and a source particle}
\label{fig:tetrahedron}
\end{figure}

In the above equation, the VEV of the braiding operator $\langle
\hat{g}_{\rho\sigma}^{(2l)} \rangle$ has been calculated exactly
in the case of compact and simple Lie groups and super Lie group,
whereas $b_u^{(l)}$ is given by
\begin{equation}
    b_u^{(l)} = \sum^{r-1}_{s=1} (q^{-Q(\rho)-Q(\sigma)+Q(\lambda_s)})^l \sqrt{\frac{E_0(\lambda_s)}{(E_0(\rho)E_0(\sigma))^3}} \langle T(\phi_u,\sigma,\bar{\sigma} | \lambda_s,\rho,\bar{\rho}) \rangle ,
\end{equation}
where
\begin{equation}
    q = exp \left(\frac{2\pi i}{k+Q(\mbox{Adj})}\right) .
\end{equation}
$Q(\rho)$ is the quadratic Casimir invariant of the $\rho$
representation. $T(\phi_n,\sigma,\bar{\sigma} |
\lambda_\rho,\rho,\bar{\rho})$ is a tetrahedron operator
(Figure~\ref{fig:tetrahedron}).

Now, we will discussed the detail calculation of the VEV of
tetrahedron operator for simple Lie group

\setlength{\unitlength}{1mm}
\begin{fmffile}{tetrahedron1}
\fmfset{arrow_len}{2mm} \fmfset{arrow_ang}{20}

Starting from $u$ channel basis $H_u$, we can construct the following expansion of $s$ channel $I_s$ \\ \\
\begin{equation}
\parbox{20mm}{
\begin{fmfgraph*}(15,15)
  \fmfleft{i1,i2}\fmfright{o1,o2}
  \fmflabel{$1$}{i1}
  \fmflabel{$2$}{o1}
  \fmflabel{$\bar 3$}{i2}
  \fmflabel{$\bar 4$}{o2}
  \fmf{fermion,label=$s$}{v1,v2}
  \fmf{fermion}{i1,v1}
  \fmf{fermion}{o1,v1}
  \fmf{fermion}{v2,o2}
  \fmf{fermion}{v2,i2}
\end{fmfgraph*}
} = \sum_{u'} L_{su'}
\parbox{20mm}{
\begin{fmfgraph*}(15,15)
  \fmfleft{i1,i2}\fmfright{o1,o2}
  \fmflabel{$1$}{i1}
  \fmflabel{$2$}{o1}
  \fmflabel{$\bar 3$}{i2}
  \fmflabel{$\bar 4$}{o2}
  \fmf{fermion,label=$u'$}{v1,v2}
  \fmf{fermion}{i1,v1}
  \fmf{fermion}{v1,i2}
  \fmf{fermion}{o1,v2,o2}
\end{fmfgraph*}
}  .
\end{equation} \\ \\
If we attach the Wilson line
\parbox{15mm}{
\begin{fmfgraph*}(10,10)
  \fmftop{t}\fmfbottom{b}\fmfright{o}
  \fmflabel{$u$}{o}
  \fmf{fermion,label=$\bar 4$}{t,v}
  \fmf{fermion,label=$2$}{v,b}
  \fmf{fermion}{v,o}
\end{fmfgraph*}
}
to the top side of both $H_u$ and $I_s$ we will get the following relation\footnote{Here, we also use the following notation $i \equiv \rho_i$ and $\bar{i} \equiv \bar{\rho}_i (i = 1, 2, 3, 4)$ as representations and its conjugate of the gauge group $G$ attach to the Wilson lines.}  \\ \\
\begin{equation}
\parbox{25mm}{
\begin{fmfgraph*}(20,20)
  \fmfleft{i1,i2}\fmfright{o1,o2}
  \fmf{fermion,label=$s$}{v1,v2}
  \fmf{fermion}{i1,v1}
  \fmf{phantom}{o1,v1}
  \fmf{phantom}{v2,o2}
  \fmf{fermion}{v2,i2}
  \fmffreeze
  \fmf{fermion,left=.7,label=$\bar 4$}{v2,v3}
  \fmf{fermion,left=.7,label=$2$}{v3,v1}
  \fmf{fermion,left=0.7,label=$\bar 3$}{i2,v4}
  \fmf{fermion,left=0.7,label=$1$}{v4,i1}
  \fmf{fermion,label=$u$}{v3,v4}
  \fmfforce{(w,0.5h)}{v4}
  \fmfforce{(0.7w,0.5h)}{v3}
  \fmfshift{(-5mm,0)}{o1,o2,i1,i2,v1,v2}
\end{fmfgraph*}
} = \sum_{u'} L_{su'} \quad
\parbox{25mm}{
\begin{fmfgraph*}(20,20)
  \fmfleft{i1,i2}\fmfright{o1,o2}
  \fmf{fermion,label=$u'$}{v1,v2}
  \fmf{fermion}{i1,v1}
  \fmf{fermion}{v1,i2}
  \fmf{phantom}{o1,v2,o2}
  \fmffreeze
  \fmf{fermion,left=0.7,label=$\bar 3$}{i2,v4}
  \fmf{fermion,left=0.7,label=$1$}{v4,i1}
  \fmf{fermion,left=1.5,label=$\bar 4$}{v2,v3}
  \fmf{fermion,left=1.5,label=$2$}{v3,v2}
  \fmfforce{(0.7w,0.5h)}{v3}
  \fmfforce{(w,0.5h)}{v4}
  \fmfshift{(-5mm,0)}{i1,i2,v1,v2,o1,o2}
  \fmf{fermion,label=$u$}{v3,v4}
\end{fmfgraph*}
} .
\end{equation} \\ \\ \\
The loop in the Wilson line can be expressed as a single Wilson line according to the following identity \\ \\
\begin{equation}
\parbox{20mm}{
\begin{fmfgraph*}(15,15)
  \fmfbottom{i}\fmftop{o}
  \fmflabel{$u'$}{i}
  \fmflabel{$u$}{o}
  \fmf{fermion,tension=0.3,left,label=$\bar 4$}{v1,v2}
  \fmf{fermion,tension=0.3,left,label=$2$}{v2,v1}
  \fmf{fermion}{i,v1}
  \fmf{fermion}{v2,o}
\end{fmfgraph*}
} = \delta_{uu'} (U^{u'}_{\bar{4}2})^{-1}
\parbox{5mm}{
\begin{fmfgraph*}(5,15)
  \fmfbottom{i}\fmftop{o}
  \fmf{fermion,label=$u'$}{i,o}
\end{fmfgraph*}
} \label{loopeqn}
\end{equation} \\ \\
then, it can be shown easily that a tetrahedron diagram can be constructed from baryon type diagrams as given in the following relation \\
\begin{eqnarray}
\parbox{25mm}{
\begin{fmfgraph*}(20,20)
  \fmfkeep{tetrahedron}
  \fmfsurround{i1,i3,i2}
  \fmf{fermion,left=0.55,label=$1$}{i1,i2}
  \fmf{fermion,left=0.55,label=$s$}{i2,i3}
  \fmf{fermion,left=0.6,label=$\bar 3$}{i3,i1}
  \fmf{fermion,label=$u$}{v,i1}
  \fmf{fermion,label=$2$}{v,i2}
  \fmf{fermion,label=$4$}{v,i3}
\end{fmfgraph*}
} &=& \sum_{u'} L_{su'}\delta_{uu'}(U^{u'}_{\bar{4}2})^{-1} \quad
\parbox{25mm}{
\begin{fmfgraph*}(20,20)
  \fmfleft{i}\fmfright{o}
  \fmf{fermion,left,label=$\bar 3$}{i,o}
  \fmf{fermion,right,label=$\bar 1$}{i,o}
  \fmf{fermion,label=$u'$}{i,o}
\end{fmfgraph*}
}, \\\nonumber \\\nonumber &=& L_{su}(U^{u}_{\bar{4}2})^{-1}
\quad\quad
\parbox{25mm}{
\begin{fmfgraph*}(20,20)
  \fmfbottom{i}\fmftop{o}
  \fmf{fermion,left,label=$\bar 3$}{i,o}
  \fmf{fermion,left,label=$1$}{o,i}
  \fmf{fermion,label=$u'$}{i,o}
\end{fmfgraph*}
}   .
\end{eqnarray}

Note that $U_{ij}^s$ is determined by connecting both endpoints in
Eq.~(\ref{loopeqn}), and therefore can be written as
\begin{equation}
U^s_{ij}=\sqrt{\frac{E_0(\rho_s)}{E_0(\rho_i)E_0(\rho_j)}} .
\end{equation}
Substituting this into the above tetrahedron-baryon relation, we get \\
\begin{equation}
\left\langle\quad\parbox{23mm}{
\begin{fmfgraph*}(20,20)
  \fmfsurround{i1,i3,i2}
  \fmf{fermion,left=0.55,label=$\bar 3$}{i1,i2}
  \fmf{fermion,left=0.55,label=$1$}{i2,i3}
  \fmf{fermion,left=0.6,label=$s$}{i3,i1}
  \fmf{fermion,label=$4$}{v,i1}
  \fmf{fermion,label=$u$}{v,i2}
  \fmf{fermion,label=$2$}{v,i3}
\end{fmfgraph*}
}\right\rangle = L_{su}\left(\frac{E_0(\bar
4)E_0(2)}{E_0(u)}\right)^{1/2} \quad
\left\langle\quad\parbox{23mm}{
\begin{fmfgraph*}(20,20)
  \fmfleft{i}\fmfright{o}
  \fmf{fermion,left,label=$\bar 3$}{i,o}
  \fmf{fermion,right,label=$\bar 1$}{i,o}
  \fmf{fermion,label=$u$}{i,o}
\end{fmfgraph*}
}\right\rangle
\end{equation} \\
which, by substituting the definition in equation
(\ref{vev-baryon}),
we finally get the VEV of tetrahedron diagram
\begin{eqnarray}
\left\langle\quad\parbox{23mm}{ \fmfreuse{tetrahedron}
}\right\rangle
&=& L_{su}\left(\frac{E_0(\bar 4)E_0(2)}{E_0(u)}\right)^{1/2}(E_0(\bar 3)E_0(u)E_0(\bar 1))^{1/2} , \\
&=& L_{su}(E_0(\bar 4)E_0(\bar 3)E_0(2)E_0(\bar 1))^{1/2},
\end{eqnarray}
or
\begin{equation}
L_{su}=(E_0(\bar 4)E_0(\bar 3)E_0(\bar 1)E_0(2))^{-1/2}
\left\langle\quad\parbox{23mm}{ \fmfreuse{tetrahedron}
}\right\rangle .
\end{equation} \\
\end{fmffile}

\begin{fmffile}{tetrahedron2}
\fmfset{arrow_len}{2mm} \fmfset{arrow_ang}{20}

One can also easily see the cyclic symmetry modulo 4 of the tetrahedron diagram as follows \\ \\
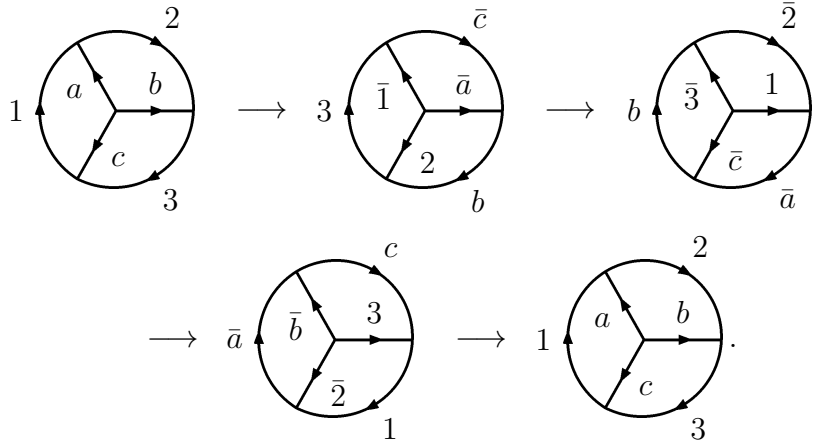
\begin{center}
\parbox{20mm}{
\begin{fmfgraph*}(20,20)
  \fmfsurround{i1,i3,i2}
  \fmf{fermion,left=0.55,label=$1$}{i2,i3}
  \fmf{fermion,left=0.6,label=$2$}{i3,i1}
  \fmf{fermion,left=0.55,label=$3$}{i1,i2}
  \fmf{fermion,label=$a$}{v,i3}
  \fmf{fermion,label=$b$}{v,i1}
  \fmf{fermion,label=$c$}{v,i2}
\end{fmfgraph*}
} \quad$\longrightarrow$\quad\quad
\parbox{20mm}{
\begin{fmfgraph*}(20,20)
  \fmfsurround{i1,i3,i2}
  \fmf{fermion,left=0.55,label=$3$}{i2,i3}
  \fmf{fermion,left=0.6,label=$\bar c$}{i3,i1}
  \fmf{fermion,left=0.55,label=$b$}{i1,i2}
  \fmf{fermion,label=$\bar 1$}{v,i3}
  \fmf{fermion,label=$\bar a$}{v,i1}
  \fmf{fermion,label=$2$}{v,i2}
\end{fmfgraph*}
} \quad$\longrightarrow$\quad\quad
\parbox{20mm}{
\begin{fmfgraph*}(20,20)
  \fmfsurround{i1,i3,i2}
  \fmf{fermion,left=0.55,label=$b$}{i2,i3}
  \fmf{fermion,left=0.6,label=$\bar 2$}{i3,i1}
  \fmf{fermion,left=0.55,label=$\bar a$}{i1,i2}
  \fmf{fermion,label=$\bar 3$}{v,i3}
  \fmf{fermion,label=$1$}{v,i1}
  \fmf{fermion,label=$\bar c$}{v,i2}
\end{fmfgraph*}
}

\vspace{10mm} \quad$\longrightarrow$\quad\quad
\parbox{20mm}{
\begin{fmfgraph*}(20,20)
  \fmfsurround{i1,i3,i2}
  \fmf{fermion,left=0.55,label=$\bar a$}{i2,i3}
  \fmf{fermion,left=0.6,label=$c$}{i3,i1}
  \fmf{fermion,left=0.55,label=$1$}{i1,i2}
  \fmf{fermion,label=$\bar b$}{v,i3}
  \fmf{fermion,label=$3$}{v,i1}
  \fmf{fermion,label=$\bar 2$}{v,i2}
\end{fmfgraph*}
} \quad$\longrightarrow$\quad\quad
\parbox{20mm}{
\begin{fmfgraph*}(20,20)
  \fmfsurround{i1,i3,i2}
  \fmf{fermion,left=0.55,label=$1$}{i2,i3}
  \fmf{fermion,left=0.6,label=$2$}{i3,i1}
  \fmf{fermion,left=0.55,label=$3$}{i1,i2}
  \fmf{fermion,label=$a$}{v,i3}
  \fmf{fermion,label=$b$}{v,i1}
  \fmf{fermion,label=$c$}{v,i2}
\end{fmfgraph*}
}\:.
\end{center}
\end{fmffile}

\vspace{15mm}



We can also define similar matrices that relates different
channels
\begin{eqnarray}
J_t = \sum_u \phi_t M_{tu} \phi^{-1}_u H_u H_u = \sum_t
\phi^{-1}_t M_{tu} \phi_u   J_t
\end{eqnarray}
and
\begin{eqnarray}
J_t = \sum_s N_{ts} \phi_s I_s I_s = \sum_t N_{ts} \phi^{-1}_s
J_t,
\end{eqnarray}
where the $\phi$ factors are defined as
\begin{eqnarray}
\phi_s = \beta_s q^{\frac{1}{2}(Q(3)+Q(4)-Q(s)}, \\
\phi_t = \beta_t q^{\frac{1}{2}(Q(t)+Q(4)-Q(1)}, \\
\phi_u = \beta_u q^{\frac{1}{2}(Q(2)+Q(4)-Q(u)},
\end{eqnarray}
with symmetry factors
\begin{eqnarray}
\beta_a=\left\{\begin{array}{cl} +1 & \mbox{{\it a} is symmetric combination of two representations} \\
-1 & \mbox{{\it a} is antisymmetric combination of two
representations}. \end{array}\right.
\end{eqnarray}

These $M$ and $N$ matrices can be found by the same manner as the
previous derivation of the $L$ matrix. In summary, the matrices
$L,M$, and $N$ are given by
\begin{eqnarray}
L_{su}&=&(E_0(1)E_0(2)E_0(3)E_0(4))^{-\frac{1}{2}} \langle T(4,s,2|1,u,\bar 3)\rangle, \\
M_{tu}&=&(E_0(1)E_0(2)E_0(3)E_0(4))^{-\frac{1}{2}} \langle T(4,u,2|3,t,1)\rangle, \\
N_{ts}&=&(E_0(1)E_0(2)E_0(3)E_0(4))^{-\frac{1}{2}} \langle
T(1,t,4|3,s,2)\rangle  .
\end{eqnarray}
From the orthonormality condition of the bases, these matrices
will satisfy the following unitary contraints
\begin{equation}
LL^\dagger=1, MM^\dagger=1, NN^\dagger=1,
\end{equation}
as well as the following orthogonality conditions, obtained from
previous relations
\begin{equation}
LL^T=1, MM^T=1, NN^T=1 .
\end{equation}
Also, from the consistency condition, one can easily derive the
following relation among these matrices
\begin{equation}
N \phi_s L = \phi_t M \phi^{-1}_u.
\end{equation}
These constraints and conditions will fix the components of these
$L, M$, and $N$ matrices.

Now, for example, if we choose $\bar{\rho}=\sigma=\underline{N}$
and $\bar{\sigma} = \rho = \overline{N}$, where $\underline{N}$
and $\overline{N}$ are the fundamental representation of $SU(N)$,
the VEV of tetrahedron operator gives
\begin{eqnarray}
    L_{id,S}&=&\frac{1}{E_0^2(N)}\langle T(S,\underline{N},\overline{N} | id,\underline{N},\underline{N}) \rangle = \frac{\sqrt{E_0(S)}}{E_0(N)} , \\
    L_{id,A}&=&\frac{1}{E_0^2(N)}\langle T(A,\underline{N},\overline{N} | id,\underline{N},\underline{N}) \rangle = \frac{\sqrt{E_0(A)}}{E_0(N)} , \\
    N_{Adj,id}&=&\frac{1}{E_0^2(N)}\langle T(id,\overline{N},\underline{N} | \mbox{Adj},\underline{N},\underline{N}) \rangle = \frac{\sqrt{E_0(\mbox{Adj})}}{E_0(N)} , \\
    N_{id,id}&=&\frac{1}{E_0^2(N)}\langle T(id,\overline{N},\underline{N} | id,\underline{N},\underline{N}) \rangle = \frac{1}{E_0(N)} .
\end{eqnarray}
The above relation are consistent with the VEV of baryon type
operator in equation $\,(\ref{vev-baryon})$

\section{Perturbative Case}
In the previous chapter, we discussed the calculation of vacuum
expectation value of extended Wilson loop operators by using
unperturbative method. This concept has been used for many
applications in various physical conditions \cite{Hahn2004,Tohru}.
Now, we will calculate the vacuum expectation value of unknotted
Wilson loop operator $\left\langle
W_{\rho}(\circlearrowleft)\right\rangle$ within the framework of
perturbation theory. The perturbative problem has been discussed
in many papers for many different settings
\cite{AxelrodSinger1991,AltschulerFreidel1997,BlasiCollina1990,Chair2007,al,alp,barnatanth}.
In this section, we use arbitrary gauge group and restrict the
order of the calculation of  $\left\langle
W_{\rho}(\circlearrowleft)\right\rangle$ up to order
$(\frac{2\pi}{k})^3$. Particularly, we compare the result with the
nonperturbative one for $SU(N)$ and $E_6$ gauge groups.

In the perturbative case, the classical CSW action
$\,(\ref{scsw})$ must undergo a modification in order to perform
the quantization. We will adopt the standard Faddeev-Popov
procedure, and the total action becomes
\cite{GuadagniniMartelliniMintchev1989}
\begin{eqnarray}
\label{stot} S_{tot}&=&S_{CS}+S_{gauge-fixing} +
S_{ghost}\nonumber\\
&=&\frac{k}{4\pi}\int_{M^3}d^{3}x\;\epsilon^{\mu\nu\rho}\;Tr\Bigg(A_{\mu}\partial_{\nu}A_{\rho}+i\frac{2}{3}\:A_{\mu}A_{\nu}A_{\rho}\Bigg)
\nonumber\\&&+\frac{k}{4\pi}\int_{M^3}d^{3}x\:\sqrt{g}\:g^{\mu\nu}\;A_{\mu}^a\partial_{\nu}\phi^a
-\int_{M^3}d^{3}x\:\sqrt{g}\:g^{\mu\nu}\;\partial_{\nu}\bar{c}^a(D_{\nu}c)^a,
\end{eqnarray}
where $c$ and $\bar{c}^a$  are Faddeev-Popov ghosts, $\phi^a$ is
the Lagrange multiplier (auxiliary field) and
\begin{equation}
(D_{\mu}c)^a=\partial_{\mu}c^a-f^{abc}A_{\mu}^bc^c.
\end{equation}
Here, $f^{abc}$ is the structure constant of the group $G$ and
$g_{\mu\nu}$ is some metric on $M^3$. The resulting propagators
from eq.\,(\ref{stot}) are
\begin{equation}
\left\langle
A_{\mu}^{a}(x)A_{\nu}^{b}(y)\right\rangle=\frac{i}{k}\delta^{ab}\epsilon_{\mu\nu\sigma}\frac{(x-y)^\sigma}{\left|x-y\right|^3},
\end{equation}
\begin{equation}
\left\langle
A_{\mu}^{a}(x)\phi^{b}(y)\right\rangle=-\frac{i}{k}\delta^{ab}\frac{(x-y)_\mu}{\left|x-y\right|^3},
\end{equation}
\begin{equation}
\left\langle \phi^{a}(x)\phi^{b}(y)\right\rangle=0,
\end{equation}
\begin{equation}
\left\langle
c^{a}(x)\bar{c}^{b}(y)\right\rangle=-\frac{i}{4\pi}\delta^{ab}\frac{1}{\left|x-y\right|}.
\end{equation}
The Wilson loop operator in a representation $\rho$ of $G$
\cite{Witten1989a,GuadagniniMartelliniMintchev1989} is defined as
\begin{eqnarray}
W_{\rho}(C)&=&\mbox{Tr}_{\rho}(\mathcal{P}\:\exp\oint_{C}A)\nonumber\\
&=&\mbox{Tr}_{\rho}\Bigg[1+i\oint_{C} dx^{\mu} A_{\mu}(x)
-\oint_{C}dx^{\mu}\int^x dy^{\nu}A_{\nu}(y)A_{\mu}(x)\nonumber\\
&&~~~~~~~~-i\oint_{C} dx^{\mu}\int^{x} dy^{\nu}\int^{y}
dz^{\rho}A_{\rho}(z)A_{\nu}(y)A_{\mu}(x)\nonumber
\end{eqnarray}
\begin{eqnarray}
\label{wrho}
&&~~~~~~~~~~~+\oint_{C} dx^{\mu}\int^x dy^{\nu}\int^y dz^{\rho}\int^z dw^{\sigma}A_{\sigma}(w)A_{\rho}(z)A_{\nu}(y)A_{\mu}(x)\nonumber\\
&+&i\oint_{C} dx^{\mu}\int^x dy^{\nu}\int^y dz^{\rho}\int^z
dw^{\sigma}\int^w dv^{\lambda} A_{\lambda}(w)
A_{\sigma}(w)A_{\rho}(z)A_{\nu}(y)A_{\mu}(x)\nonumber\\
&-& \oint_{C} dx^{\mu}\int^x dy^{\nu}\int^y dz^{\rho}\int^z
dw^{\sigma}\int^w dv^{\lambda}\int^v du^{\tau}\nonumber\\
&&~~~~~~~~~~~~\times A_{\tau}(u)A_{\lambda}(w)
A_{\sigma}(w)A_{\rho}(z)A_{\nu}(y)A_{\mu}(x)+ . . .\Bigg].
\end{eqnarray}
All line integrals are performed on the same contour $C$. If an
explicit parameterizations $\left\{x^{\mu}(t) : 0 \leq t \leq
1\right\}$ of  $C$ is used, then we will get
\begin{equation}
\label{para1} \oint_{C}dx^{\mu}\int^x dy^{\nu}=\int^1_0
ds\:\int^s_0 dt\:\dot{x}^{\mu}(s)\dot{x}^{\nu}(t),
\end{equation}
and so on.

\subsection{The Ghost Fields Contribution}
From eq.$\,(\ref{stot})$, the total quantized action contains
contributions from gauge, ghost and auxiliary fields. This can be
rewritten again as
\begin{eqnarray}
\label{scs} S_{tot}&=&\int
d^{3}x\:\Bigg(\frac{k}{4\pi}\:c_2(\rho)\:\epsilon^{\mu\nu\rho}\:A^a_{\mu}\partial_{\nu}A^a_{\rho}+\frac{k}{4\pi}A^a_{\mu}\partial^{\mu}\phi^a+\bar{c}^a\partial^{\mu}\partial_{\mu}c^a\Bigg)\nonumber\\
&&+\int
d^{3}x\:f^{abc}\:\Bigg(\partial^{\mu}\bar{c}^a\:A^b_{\mu}c^c-\frac{k}{12\pi}\:c_2(\rho)\:\epsilon^{\mu\nu\rho}\:A^a_{\mu}A^b_{\nu}A^c_{\rho}\Bigg),
\end{eqnarray}
where $c_2(\rho)$ is the quadratic Casimir for the fundamental
representation and dim $\rho$ is the dimension of the gauge group.

The standard VEV of the Wilson loop operator of the action
$\,(\ref{scs})$ can be written as
\begin{equation}
\label{defwrho} \langle
W_{\rho}(C)\rangle=\int\mathcal{D}A\:\mathcal{D}\phi\:\mathcal{D}c\:\mathcal{D}\bar{c}\:\mbox{Tr}_{\rho}\Bigg(\mathcal{P}\:\exp\oint_{C}A
\Bigg)\:e^{iS_{tot}}.
\end{equation}
To solve eq. $\,(\ref{defwrho})$ perturbatively, we insert
external source functions $J$ and $H$ in the partition function
for gauge and ghost fields respectively, and the VEV of the Wilson
loop operators is obtained as follows
\begin{eqnarray}
\langle
W_{\rho}(C)\rangle&=&\Bigg[\dim\:\rho+c_2(\rho)\oint_{C}dx^{\tau}\int^x
dy^{\lambda}\frac{\delta^2}{\delta J^{i\tau}\delta
J^{i\lambda}}\nonumber\\&&~~~~~+\frac{i}{2}\:c_2(\rho)\:f^{ijk}\oint_{C}dx^{\tau}\int^x
dy^{\lambda}\int^y dz^{\rho}\frac{\delta^3}{\delta J^{i\tau}\delta
J^{j\lambda}\delta J^{k\rho}}+...\Bigg]\nonumber
\end{eqnarray}
\begin{eqnarray}
\label{wrhoghost1}
&&\times \exp\Bigg[-f^{def}\int d^3
x\Bigg(\partial^{\sigma}\frac{1}{\delta
H^d}\Bigg)\frac{\delta^3}{\delta J^{e\sigma}\delta{\bar
H}^f}\Bigg]\nonumber\\&&\times
\exp\Bigg[\frac{k}{12\pi}\:c_2(\rho)\:\int d^3
x\:f^{abc}\epsilon^{\alpha\beta\gamma}\frac{\delta^3}{\delta
J^{a\alpha}\delta J^{b\beta}\delta
J^{c\gamma}}\Bigg]Z_0\Bigg|_{J=H=\bar{H}=0},
\end{eqnarray}
where $Z_0$ is the partition functional that is defined as
\begin{eqnarray}
Z_0&=& \exp\Bigg[\int d^3 x\:d^3
y\Bigg(\frac{1}{8\:c_2(\rho)}J^s_{\mu}(x)\:V^{st,\mu\nu}_{AA}(x-y)\:J^t_{\nu}(y)\nonumber\\&&~~~~~~~~~~~~~+H^s(x)\:V^{st}_{\bar{c}c}(x-y)\:\bar{H}^t(y)+F(V_{A\phi})\Bigg)\Bigg],
\end{eqnarray}
where $V^{ab,\mu\nu}_{AA}(x-y)=\left\langle
A_{\mu}^{a}(x)A_{\nu}^{b}(y)\right\rangle$ and
$V^{ab}_{\bar{c}c}(x-y)=\left\langle
c^{a}(x)\bar{c}^{b}(y)\right\rangle$. Note that the contribution
of auxiliary fields vanish, because the total action
$\,(\ref{stot})$ does not contain any vertex of the auxiliary
fields.

One would expect that the ghost contribution vanish for all orders
due to the unphysical nature of the ghost fields. We show that
this contributions vanish up to the second order.

If eq. $(\ref{wrhoghost1})$ is expanded as power series of
$(1/k)$, we get

(i) For order $(1/k)^0$, $\langle W_{\rho}(C)\rangle = \dim
\:\rho$.

(ii) For order $(1/k)$, the VEV of the Wilson loop operator for
ghost fields is defined as

\begin{eqnarray}
\label{wrhoghost2} \langle
W_{\rho}(C)\rangle^{(1)}_{ghost}&=&\frac{\dim\:
\rho}{2!}\Bigg[\int d^3
x\:f^{def}\Bigg(\partial^{\sigma}\frac{1}{\delta
H^d}\Bigg)\frac{\delta^3}{\delta J^{e\sigma}\delta{\bar
H}^f}\Bigg]^2\:Z_0\Bigg|_{J=H=\bar{H}=0}\nonumber\\
&&-\dim\: \rho\Bigg[\int d^3
x\:f^{def}\Bigg(\partial^{\sigma}\frac{1}{\delta
H^d}\Bigg)\frac{\delta^3}{\delta J^{e\sigma}\delta{\bar
H}^f}\Bigg]\nonumber\\&&~\times\Bigg[\frac{k\:c_2(\rho)}{12\pi}\:\int
d^3 x\:f^{abc}\epsilon^{\alpha\beta\gamma}\frac{\delta^3}{\delta
J^{a\alpha}\delta J^{b\beta}\delta
J^{c\gamma}}\Bigg]Z_0\Bigg|_{J=H=\bar{H}=0}.
\end{eqnarray}
One can see that the value of $\langle
W_{\rho}(C)\rangle^{(1)}_{ghost}$ vanish.

(iii) For order $(1/k)^2$, the VEV of the Wilson loop operator for
ghost fields contain seven terms that can be written as
\begin{eqnarray}
\langle
W_{\rho}(C)\rangle^{(2)}_{ghost}&=&-\frac{i}{2}f^{ijk}\:c_2(\rho)\:\oint_{C}dx^{\tau}\int^x
dy^{\lambda}\int^y dz^{\rho}\frac{\delta^3}{\delta J^{i\tau}\delta
J^{j\lambda}\delta J^{k\rho}}\nonumber\\
&&~~~~\times\Bigg[\int d^3
x\:f^{def}\Bigg(\partial^{\sigma}\frac{1}{\delta
H^d}\Bigg)\frac{\delta^3}{\delta J^{e\sigma}\delta{\bar
H}^f}\Bigg]\:Z_0\Bigg|_{J=H=\bar{H}=0}\nonumber
\end{eqnarray}
\begin{eqnarray}
\label{wrhoghost3} &&-c_2(\rho)\:\oint_{C}dx^{\tau}\int^x
dy^{\lambda}\frac{\delta^2}{\delta J^{i\tau}\delta
J^{i\lambda}}\Bigg[\int d^3
x\:f^{def}\Bigg(\partial^{\sigma}\frac{1}{\delta
H^d}\Bigg)\frac{\delta^3}{\delta J^{e\sigma}\delta{\bar
H}^f}\Bigg]\nonumber\\
&&~\times\Bigg[\frac{k}{12\pi}\:c_2(\rho)\:\int d^3
x\:f^{abc}\epsilon^{\alpha\beta\gamma}\frac{\delta^3}{\delta
J^{a\alpha}\delta J^{b\beta}\delta
J^{c\gamma}}\Bigg]Z_0\Bigg|_{J=H=\bar{H}=0}\nonumber\\
&&-\frac{\dim\:\rho}{3!}\Bigg[\int d^3
x\:f^{def}\Bigg(\partial^{\sigma}\frac{1}{\delta
H^d}\Bigg)\frac{\delta^3}{\delta J^{e\sigma}\delta{\bar
H}^f}\Bigg]\nonumber\\
&&~\times\Bigg[\frac{k}{12\pi}\:c_2(\rho)\int d^3
x\:f^{abc}\epsilon^{\alpha\beta\gamma}\frac{\delta^3}{\delta
J^{a\alpha}\delta J^{b\beta}\delta
J^{c\gamma}}\Bigg]^3\:Z_0\Bigg|_{J=H=\bar{H}=0}\nonumber\\
&&+\frac{\dim\: \rho}{2!}\Bigg[\int d^3
x\:f^{def}\Bigg(\partial^{\sigma}\frac{1}{\delta
H^d}\Bigg)\frac{\delta^3}{\delta J^{e\sigma}\delta{\bar
H}^f}\Bigg]^2\nonumber\\
&&~\times\frac{1}{2!}\Bigg[\frac{k}{12\pi}\:c_2(\rho)\int d^3
x\:f^{abc}\epsilon^{\alpha\beta\gamma}\frac{\delta^3}{\delta
J^{a\alpha}\delta J^{b\beta}\delta
J^{c\gamma}}\Bigg]^2\:Z_0\Bigg|_{J=H=\bar{H}=0}\nonumber\\
&&+\frac{\dim\:\rho}{4!}\Bigg[\int d^3
x\:f^{def}\Bigg(\partial^{\sigma}\frac{1}{\delta
H^d}\Bigg)\frac{\delta^3}{\delta J^{e\sigma}\delta{\bar
H}^f}\Bigg]^4\:Z_0\Bigg|_{J=H=\bar{H}=0}\nonumber\\
&&+\frac{1}{2!}\:c_2(\rho)\oint_{C}dx^{\tau}\int^x
dy^{\lambda}\frac{\delta^2}{\delta J^{i\tau}\delta
J^{i\lambda}}\nonumber\\&&~~\times\Bigg[\int d^3
x\:f^{def}\Bigg(\partial^{\sigma}\frac{1}{\delta
H^d}\Bigg)\frac{\delta^3}{\delta J^{e\sigma}\delta{\bar
H}^f}\Bigg]^2\:Z_0\Bigg|_{J=H=\bar{H}=0}\nonumber\\
&&-\frac{\dim\:\rho}{3!}\Bigg[\int d^3
x\:f^{def}\Bigg(\partial^{\sigma}\frac{1}{\delta
H^d}\Bigg)\frac{\delta^3}{\delta J^{e\sigma}\delta{\bar
H}^f}\Bigg]^3\nonumber\\
&&~\times\Bigg[\frac{k}{12\pi}\:c_2(\rho)\int d^3
x\:f^{abc}\epsilon^{\alpha\beta\gamma}\frac{\delta^3}{\delta
J^{a\alpha}\delta J^{b\beta}\delta
J^{c\gamma}}\Bigg]Z_0\Bigg|_{J=H=\bar{H}=0}.
\end{eqnarray}
Because of the anticommutation and the peculiar structure of the
indices (see Appendix A), eq.$\,(\ref{wrhoghost3})$ become simpler
\begin{eqnarray}
\langle W_{\rho}(C)\rangle^{(2)}_{ghost}&=&
\frac{\dim\:\rho}{4!}\Bigg[\int d^3
x\:f^{def}\Bigg(\partial^{\sigma}\frac{1}{\delta
H^d}\Bigg)\frac{\delta^3}{\delta J^{e\sigma}\delta{\bar
H}^f}\Bigg]^4\:Z_0\Bigg|_{J=H=\bar{H}=0}\nonumber\\
&&+\frac{1}{2!}\:c_2(\rho)\oint_{C}dx^{\tau}\int^x
dy^{\lambda}\frac{\delta^2}{\delta J^{i\tau}\delta
J^{i\lambda}}\nonumber\\&&~\times\Bigg[\int d^3
x\:f^{def}\Bigg(\partial^{\sigma}\frac{1}{\delta
H^d}\Bigg)\frac{\delta^3}{\delta J^{e\sigma}\delta{\bar
H}^f}\Bigg]^2\:Z_0\Bigg|_{J=H=\bar{H}=0}\nonumber
\end{eqnarray}
\begin{eqnarray}
\label{wrhoghost4} &&-\frac{\dim\:\rho}{3!}\Bigg[\int d^3
x\:f^{def}\Bigg(\partial^{\sigma}\frac{1}{\delta
H^d}\Bigg)\frac{\delta^3}{\delta J^{e\sigma}\delta{\bar
H}^f}\Bigg]^3\nonumber\\
&&~\times\Bigg[\frac{k}{12\pi}\:c_2(\rho)\int d^3
x\:f^{abc}\epsilon^{\alpha\beta\gamma}\frac{\delta^3}{\delta
J^{a\alpha}\delta J^{b\beta}\delta
J^{c\gamma}}\Bigg]Z_0\Bigg|_{J=H=\bar{H}=0}.
\end{eqnarray}
Finally, eq. $\,(\ref{wrhoghost4})$ vanish because of various
reasons that will be explained in Appendix A.

\subsection{The Zeroth, First and Second Order Contribution of Gauge Fields}
In this section, we calculate perturbatively the zeroth, first,
and second order contributions of gauge fields following
Guadagnini et.al.\cite{GuadagniniMartelliniMintchev1989}. In the
next section we extend the calculation up to the third order
contribution. In order to simplify the calculation, the unknotted
knot is chosen as a circle
\begin{equation}
\label{uo} U_0=\left\{x(s)=(\cos 2\pi s, \sin 2\pi s,0);~~0\leq s
\leq 1\right\}.
\end{equation}
For the $(\frac{2\pi}{k})^0$ contribution to $\left\langle
W_{\rho}(C)\right\rangle$, we will get
\begin{equation}
\label{dim} \left\langle
W_{\rho}(C)\right\rangle^{(0)}=\left\langle
W_{\rho}(\circlearrowleft)\right\rangle^{(0)}=\dim\:\rho.
\end{equation}
Then, $(\frac{2\pi}{k})$ contribution to $\left\langle
W_{\rho}(C)\right\rangle^{(1)}$ is defined as
\begin{eqnarray}
\label{wc1} \left\langle
W_{\rho}(C)\right\rangle^{(1)}&=&-\mbox{Tr}(R^b R^a)\oint_{C}
dx^{\mu}\int^x dy^{\nu}\left\langle
A_{\nu}^b(y)A_{\mu}^a(x)\right\rangle\nonumber\\
&=&-i\left(\frac{2\pi}{k}\right)\dim\: \rho
\:c_2(\rho)\:\varphi(C),
\end{eqnarray}
where the quadratic Casimir for the fundamental representation
$c_{2}(\rho)$ is given by
\begin{equation}
c_2(\rho)1 = R^aR^a,
\end{equation}
and $\varphi(C)$ is defined as
\begin{equation}
\label{ordeone} \varphi(C)=\frac{1}{2\pi}\int^1_0 ds\:\int^s_0
dt\:\epsilon_{\mu\nu\sigma}\dot{x}^{\mu}(s)\dot{x}^{\nu}(t)
\frac{(x(s)-x(t))^{\sigma}}{\left|x(s)-x(t)\right|^3}.
\end{equation}

The formula $\,(\ref{ordeone})$ is known as the cotorsion of $C$.
The cotorsion is not invariant under the deformation of $C$,
because it is metric dependent. This is contrary to the fact that
the $\left\langle W_{\rho}(C)\right\rangle$ is a topological
invariant.  This problem can be solved by inserting a framing
contour $C_f$ that is defined as
\begin{equation}
\label{fram} x^{\mu}\rightarrow y^{\mu}=x^{\mu}+\epsilon
n^{\mu}(t)~~~~~~~,~~~~~~~\left(\epsilon >0,|n(t)|=1\right),
\end{equation}
where $n^{\mu}$ is a vector field orthogonal to $C$. In this
paper, we choose the value of $n^{\mu}$ to be
\begin{equation}
\label{ns} n(s)=[0,0,e^{\pi i s}].
\end{equation}
If the formula $\,(\ref{wc1})$ is rewritten by inserting a framing
contour $C_f$ $\,(\ref{fram})$ for the unknot
condition$\,(\ref{uo})$, we obtain
\begin{equation}
\label{wcuo} \left\langle
W_{\rho}(\circlearrowleft)\right\rangle^{(1)}_{f}=-i\left(\frac{2\pi}{k}\right)\dim\:
\rho\:c_2(\rho)\varphi_{f}(U_0)=0,
\end{equation}
where $\varphi_{f}(U_0)$ is the value of $\varphi(C)$ with
inserted framing contour $C_f$ $\,(\ref{fram})$ for the unknot
$\,(\ref{uo})$.

Now, we will analyze the  $(\frac{2\pi}{k})^{2}$ contribution to
$\left\langle W_{\rho}(C)\right\rangle$ which results from the
interactions part of the Lagrangian contributed by the $A^3$ and
$A^4$ terms of eq. $\,(\ref{wrho})$.

The first term of the $(\frac{2\pi}{k})^{2}$ part of $\left\langle
W_{\rho}(C)\right\rangle$ can be written as
\begin{eqnarray}
\label{orde2a}
\left\langle W_{\rho}(C)\right\rangle^{(2a)}&=&\mbox{Tr}_{\rho}\left[-i\oint_{C}dx^{\mu}\int^x dy^{\nu}\int^y dz^{\rho}\left\langle A_{\rho}(z)A_{\nu}(y)A_{\mu}(x)\right\rangle\right]\nonumber\\
&=&\left(\frac{2\pi}{k}\right)^2 \Bigg(\frac{\dim\:\rho\:c_v
\:c_2(\rho)}{32\pi^3}\Bigg)\times\nonumber\\
&&~~~~~~~~\times\oint_{C} dx^{\mu}\int^x dy^{\nu}\int^y
dz^{\rho}\:H_{\mu\nu\rho}(x,y,z),
\end{eqnarray}
where the quadratic Casimir for the adjoint representation $c_v$
is obtained through the relation
\begin{equation}
\label{cv} \delta^{ab}\: c_v = f^{acd}\:f^{bcd},
\end{equation}
and
\begin{equation}
\label{H}
H_{\mu\nu\rho}(x,y,z)=\epsilon^{\alpha\beta\gamma}\epsilon_{\mu\alpha\sigma}\epsilon_{\nu\beta\lambda}\epsilon_{\rho\gamma\tau}\int
d^3l\frac{(l-x)^\sigma}{\left|l-x\right|^3}
\frac{(l-y)^\lambda}{\left|l-y\right|^3}\frac{(l-z)^\tau}{\left|l-z\right|^3}.
\end{equation}
If we use the unknotted knot $(\ref{uo})$ in equation $(\ref{H})$,
we will obtain
\begin{eqnarray}
\label{z1uo} \zeta_1(U_0)&=&\frac{1}{32\pi^3}\oint_{C}
dx^{\mu}\int^x dy^{\nu}\int^y
dz^{\rho}H_{\mu\nu\rho}(x,y,z)\nonumber\\
&=&-\frac{1}{16\pi^3}\int_0^{2\pi}
d\theta\int_0^{\theta}d\phi\int_0^{\phi} d\psi\times\nonumber\\&&
~~~~~~~~\times\Bigg[\sin\left(\frac{\theta -
\phi}{2}\right)+\sin\left(\frac{\theta -
\psi}{2}\right)+\sin\left(\frac{\phi -
\psi}{2}\right)\Bigg]\nonumber\\
&=&-\frac{1}{12},
\end{eqnarray}
and we get the value of $\left\langle
W_{\rho}(\circlearrowleft)\right\rangle^{(2a)}$ as
\begin{equation}
\label{orde2auo} \left\langle
W_{\rho}(\circlearrowleft)\right\rangle^{(2a)}=-\frac{1}{12}\left(\frac{2\pi}{k}\right)^2
\dim\:\rho\:c_v \:c_2(\rho).
\end{equation}
The second term of the $(\frac{2\pi}{k})^{2}$ part of
$\left\langle W_{\rho}(C)\right\rangle$ can be written as
\cite{GuadagniniMartelliniMintchev1989}
\begin{eqnarray}
\label{orde2b} \left\langle
W_{\rho}(C)\right\rangle^{(2b)}&=&\mbox{Tr}_{\rho}\left[\oint_{C}dx^{\mu}\int^x
dy^{\nu}\int^y dz^{\rho}\int^z dw^{\sigma}
\left\langle A_{\sigma}(w)A_{\rho}(z)A_{\nu}(y)A_{\mu}(x)\right\rangle\right]\nonumber\\
&=&-\frac{1}{2}\left(\frac{2\pi}{k}\right)^2 \dim\:
\rho\:c_2^2(\rho)\varphi^2(C)+\left(\frac{2\pi}{k}\right)^2 \dim\:
\rho\:c_v \:c_2(\rho)\zeta_2(C),
\end{eqnarray}
where $\varphi(C)$ is defined in eq.$\,(\ref{ordeone})$ and
$\zeta_2(C)$ is defined as
\begin{equation}
\zeta_2(C)=\frac{1}{8\pi^2}\oint_{C}dx^{\mu}\int^x dy^{\nu}\int^y
dz^{\rho}\int^z
dw^{\sigma}\epsilon_{\sigma\nu\alpha}\epsilon_{\rho\mu\beta}\frac{(w-y)^\alpha}{\left|w-y\right|^3}\frac{(z-x)^\beta}{\left|z-x\right|^3}.
\end{equation}
If we use the unknotted knot $(\ref{uo})$ in equation
$\,(\ref{orde2b})$, the  value of $\left\langle
W_{\rho}(\circlearrowleft)\right\rangle^{(2b)}$ is
\begin{equation}
\label{w2b}
 \left\langle W_{\rho}(\circlearrowleft)\right\rangle^{(2b)}=0,
\end{equation}
where we have applied the framing procedure similar to the first
order case $(\ref{wcuo})$. More specific calculations can be found
in \cite{GuadagniniMartelliniMintchev1989}. Note that the
equations $\,(\ref{orde2a})$ and $\,(\ref{orde2b})$ are
contributions of order $(\frac{2\pi}{k})^{2}$ to $\left\langle
W_{\rho}(C)\right\rangle$.

\subsection{The $(\frac{2\pi}{k})^3$ Contributions}
In this section, we discuss the contribution of order
$(\frac{2\pi}{k})^3$ to $\left\langle W_{\rho}(C)\right\rangle$.
It is divided into two parts, $\left\langle
W_{\rho}(C)\right\rangle^{(3a)}$ and $\left\langle
W_{\rho}(C)\right\rangle^{(3b)}$. $\left\langle
W_{\rho}(C)\right\rangle^{(3a)}$ contains the interaction part of
the Lagrangian contracted with the $A^5$ term of eq.
$\,(\ref{wrho})$, that is
\begin{eqnarray}
\left\langle
W_{\rho}(C)\right\rangle^{(3a)}&=&\mbox{Tr}\Bigg[i\oint_{C}
dx^{\mu}\int^x dy^{\nu}\int^y dz^{\rho}\int^z dw^{\sigma}\int^w
dv^{\lambda}\times\nonumber\\
&&\times
\left\langle A_{\lambda}(v)A_{\sigma}(w)A_{\rho}(z)A_{\nu}(y)A_{\mu}(x)\right\rangle\Bigg]\nonumber\\
 &=&\frac{ic_v \:\dim\:\rho\: c_2^2(\rho)}{8\pi \:k^3}\oint_C dx^{\mu}\int^x
dy^{\nu}\int^y dz^{\rho}\int^z dw^{\sigma}\int^w
dv^{\lambda}\times\nonumber\\
&&\times[F_{\lambda\sigma,\rho\nu\mu}(v-w,y,x,z)+F_{\lambda\rho,\sigma\nu\mu}(v-z,y,x,w)\nonumber\\
&&~~~+F_{\lambda\nu,\sigma\rho\mu}(v-y,z,x,w)+F_{\lambda\mu,\sigma\rho\nu}(v-x,z,y,w)\nonumber\\
&&~~~+F_{\sigma\rho,\lambda\nu\mu}(w-z,y,x,v)+F_{\sigma\nu,\lambda\rho\mu}(w-y,z,x,v)\nonumber\\
&&~~~+F_{\sigma\mu,\lambda\rho\nu}(w-x,z,y,v)+F_{\rho\nu,\lambda\sigma\mu}(z-y,w,x,v)\nonumber\\
&&~~~+F_{\rho\mu,\lambda\sigma\nu}(z-x,w,y,v)+F_{\nu\mu,\lambda\sigma\rho}(y-x,w,z,v)]\nonumber
\end{eqnarray}
\begin{eqnarray}
\label{wrho3a}
&&-\frac{ic_v^2\:\dim~\rho\:c_2(\rho)}{16\pi\:k^3}\oint_C
x^{\mu}\int^x
dy^{\nu}\int^y dz^{\rho}\int^z dw^{\sigma}\int^w dv^{\lambda}\times\nonumber\\
&&\times[F_{\lambda\rho,\sigma\nu\mu}(v-z,y,x,w)+F_{\lambda\nu,\sigma\rho\mu}(v-y,z,x,w)\nonumber\\
&&~~~+F_{\sigma\nu,\lambda\rho\mu}(w-y,z,x,v)+F_{\sigma\mu,\lambda\rho\nu}(w-x,z,y,v)\nonumber\\
&&~~~+F_{\rho\mu,\lambda\sigma\nu}(z-x,w,y,v)],
\end{eqnarray}

where $F_{\lambda\sigma,\rho\nu\mu}(v-w,y,x,z)$ is given by

\begin{equation}
F_{\lambda\sigma,\rho\nu\mu}(v-w,y,x,z)=\epsilon_{\lambda\sigma\alpha}\frac{(v-w)^{\alpha}}{\left|v-w\right|^3}H_{\rho\nu\mu}(y-z,x-z).
\end{equation}

The $\left\langle W_{\rho}(C)\right\rangle^{(3b)}$ contribution is
related to the $A^6$ term of eq. $\,(\ref{wrho})$. This
contribution is written in equation (\ref{wrho3b}) and includes
only terms of order $(\frac{2\pi}{k})^3$ which form combinations
of three gauge propagators. The $A^6$ terms that involve
combinations of two gauge vertices are excluded since they are of
order $(\frac{2\pi}{k})^4$. The contribution $\left\langle
W_{\rho}(C)\right\rangle^{(3b)}$ is defined as
\begin{eqnarray}
\left\langle
W_{\rho}(C)\right\rangle^{(3b)}&=&\mbox{Tr}\Bigg[i\oint_{C}
dx^{\mu}\int^x dy^{\nu}\int^y dz^{\rho}\int^z dw^{\sigma}\int^w
dv^{\lambda}\int^v du^{\tau}\times\nonumber\\&&~~~~~~~~~~~~\times
\left\langle
A_{\tau}(u)A_{\lambda}(v)A_{\sigma}(w)A_{\rho}(z)A_{\nu}(y)A_{\mu}(x)\right\rangle\Bigg]~~~~\nonumber\\
&=&\frac{i\:\dim\rho\:c_2^3(\rho)}{k^3}\oint_C dx^{\mu}\int^x
dy^{\nu}\int^y dz^{\rho}\int^z dw^{\sigma}\int^w
dv^{\lambda}\int^v du^{\tau}\times\nonumber\\
&&\times
\Bigg[\epsilon_{\tau\lambda\alpha}\epsilon_{\sigma\rho\beta}\epsilon_{\nu\mu\gamma}\frac{(u-v)^{\alpha}}{\left|u-v\right|^3}\frac{(w-z)^{\beta}}{\left|w-z\right|^3}\frac{(y-x)^{\gamma}}{\left|y-x\right|^3}\nonumber\\
& &+\epsilon_{\tau\lambda\alpha}\epsilon_{\sigma\nu\beta}\epsilon_{\rho\mu\gamma}\frac{(u-v)^{\alpha}}{\left|u-v\right|^3}\frac{(w-y)^{\beta}}{\left|w-y\right|^3}\frac{(z-x)^{\gamma}}{\left|z-x\right|^3}\nonumber\\
& &+\epsilon_{\tau\lambda\alpha}\epsilon_{\sigma\mu\beta}\epsilon_{\rho\nu\gamma}\frac{(u-v)^{\alpha}}{\left|u-v\right|^3}\frac{(w-x)^{\beta}}{\left|w-x\right|^3}\frac{(z-y)^{\gamma}}{\left|z-y\right|^3}\nonumber\\
&
&+\epsilon_{\tau\rho\alpha}\epsilon_{\lambda\sigma\beta}\epsilon_{\nu\mu\gamma}\frac{(u-z)^{\alpha}}{\left|u-z\right|^3}\frac{(v-w)^{\beta}}{\left|v-w\right|^3}\frac{(y-x)^{\gamma}}{\left|y-x\right|^3}\nonumber\\
& &+\Bigg(1-\frac{k_1}{2}\Bigg)\epsilon_{\tau\sigma\alpha}\epsilon_{\lambda\rho\beta}\epsilon_{\nu\mu\gamma}\frac{(u-w)^{\alpha}}{\left|u-w\right|^3}\frac{(v-z)^{\beta}}{\left|v-z\right|^3}\frac{(y-x)^{\gamma}}{\left|y-x\right|^3}\nonumber\\
&
&+\Bigg(1-k_1+\frac{k_1^2}{4}\Bigg)\epsilon_{\tau\sigma\alpha}\epsilon_{\lambda\nu\beta}\epsilon_{\rho\mu\gamma}\frac{(u-w)^{\alpha}}{\left|u-w\right|^3}\frac{(v-y)^{\beta}}{\left|v-y\right|^3}\frac{(z-x)^{\gamma}}{\left|z-x\right|^3}\nonumber
\end{eqnarray}
\begin{eqnarray}
\label{wrho3b}
&
&+\Bigg(1-\frac{k_1}{2}\Bigg)\epsilon_{\tau\sigma\alpha}\epsilon_{\lambda\mu\beta}\epsilon_{\rho\nu\gamma}\frac{(u-w)^{\alpha}}{\left|u-w\right|^3}\frac{(v-x)^{\beta}}{\left|v-x\right|^3}\frac{(z-y)^{\gamma}}{\left|z-y\right|^3}\nonumber\\
& &+\Bigg(1-\frac{3k_1}{2}+\frac{k_1^2}{2}\Bigg)\epsilon_{\tau\rho\alpha}\epsilon_{\lambda\nu\beta}\epsilon_{\sigma\mu\gamma}\frac{(u-z)^{\alpha}}{\left|u-z\right|^3}\frac{(v-y)^{\beta}}{\left|v-y\right|^3}\frac{(w-x)^{\gamma}}{\left|w-x\right|^3}\nonumber\\
& &+\Bigg(1-k_1+\frac{k_1^2}{4}\Bigg)\epsilon_{\tau\rho\alpha}\epsilon_{\lambda\mu\beta}\epsilon_{\sigma\nu\gamma}\frac{(u-z)^{\alpha}}{\left|u-z\right|^3}\frac{(v-x)^{\beta}}{\left|v-x\right|^3}\frac{(w-y)^{\gamma}}{\left|w-y\right|^3}\nonumber\\
& &+\Bigg(1-\frac{k_1}{2}\Bigg)\epsilon_{\tau\nu\alpha}\epsilon_{\lambda\sigma\beta}\epsilon_{\rho\mu\gamma}\frac{(u-y)^{\alpha}}{\left|u-y\right|^3}\frac{(v-w)^{\beta}}{\left|v-w\right|^3}\frac{(z-x)^{\gamma}}{\left|z-x\right|^3}\nonumber\\
& &+\Bigg(1-k_1+\frac{k_1^2}{4}\Bigg)\epsilon_{\tau\nu\alpha}\epsilon_{\lambda\rho\beta}\epsilon_{\sigma\mu\gamma}\frac{(u-y)^{\alpha}}{\left|u-y\right|^3}\frac{(v-z)^{\beta}}{\left|v-z\right|^3}\frac{(w-x)^{\gamma}}{\left|w-x\right|^3}\nonumber\\
& &+\Bigg(1-\frac{k_1}{2}\Bigg)\epsilon_{\tau\nu\alpha}\epsilon_{\lambda\mu\beta}\epsilon_{\sigma\rho\gamma}\frac{(u-y)^{\alpha}}{\left|u-y\right|^3}\frac{(v-x)^{\beta}}{\left|v-x\right|^3}\frac{(w-z)^{\gamma}}{\left|w-z\right|^3}\nonumber\\
& &+\epsilon_{\tau\mu\alpha}\epsilon_{\lambda\sigma\beta}\epsilon_{\rho\nu\gamma}\frac{(u-x)^{\alpha}}{\left|u-x\right|^3}\frac{(v-w)^{\beta}}{\left|v-w\right|^3}\frac{(z-y)^{\gamma}}{\left|z-y\right|^3}\nonumber\\
& &+\Bigg(1-\frac{k_1}{2}\Bigg)\epsilon_{\tau\mu\alpha}\epsilon_{\lambda\rho\beta}\epsilon_{\sigma\nu\gamma}\frac{(u-x)^{\alpha}}{\left|u-x\right|^3}\frac{(v-z)^{\beta}}{\left|v-z\right|^3}\frac{(w-y)^{\gamma}}{\left|w-y\right|^3}\nonumber\\
&
&+\epsilon_{\tau\mu\alpha}\epsilon_{\lambda\nu\beta}\epsilon_{\rho\sigma\gamma}\frac{(u-x)^{\alpha}}{\left|u-x\right|^3}\frac{(v-y)^{\beta}}{\left|v-y\right|^3}\frac{(z-w)^{\gamma}}{\left|z-w\right|^3}\Bigg],
\end{eqnarray}
where $k_1 = c_v/c_2(\rho)$. If we use the unknot condition
$\,(\ref{uo})$ in eq. $\,(\ref{wrho3a})$, we get the integral

\begin{eqnarray}
\label{int3a1}&&\oint_C dx^{\mu}\int^x dy^{\nu}\int^y
dz^{\rho}\int^z dw^{\sigma}\int^w
dv^{\lambda}\epsilon_{\rho\mu\alpha}\frac{(z-x)^{\alpha}}{\left|z-x\right|^3}H_{\lambda\sigma\nu}(w-v,y-v)\nonumber\\
&&~~~~~~~~~~~~~~~~~~~~~~~~~~~~
=32i\pi\Bigg(\frac{\pi^2}{6}-1\Bigg),
\end{eqnarray}
\begin{eqnarray}
\label{int3a2} &&\oint_C dx^{\mu}\int^x dy^{\nu}\int^y
dz^{\rho}\int^z dw^{\sigma}\int^w
dv^{\lambda}\epsilon_{\sigma\mu\alpha}\frac{(w-x)^{\alpha}}{\left|w-x\right|^3}H_{\lambda\rho\nu}(z-v,y-v)\nonumber\\
&&~~~~~~~~~~~~~~~~~~~~~~~~~~~~=32i\pi\Bigg(\frac{\pi^2}{6}-1\Bigg),
\end{eqnarray}
\begin{eqnarray}
\label{int3a3} &&\oint_C dx^{\mu}\int^x dy^{\nu}\int^y
dz^{\rho}\int^z dw^{\sigma}\int^w
dv^{\lambda}\Bigg[\epsilon_{\lambda\rho\alpha}\frac{(v-z)^{\alpha}}{\left|v-z
\right|^3}H_{\sigma\nu\mu}(y-w,x-w)\nonumber\\
&&~~~~~~~~~~~+\epsilon_{\sigma\nu\alpha}\frac{(w-y)^{\alpha}}{\left|w-y
\right|^3}H_{\lambda\rho\mu}(z-v,x-v)\Bigg]=32i\pi,
\end{eqnarray}
\begin{equation}
\label{int3a4} \oint_C dx^{\mu}\int^x dy^{\nu}\int^y
dz^{\rho}\int^z dw^{\sigma}\int^w
dv^{\lambda}\epsilon_{\lambda\nu\alpha}\frac{(v-y)^{\alpha}}{\left|v-y\right|^3}H_{\sigma\rho\mu}(z-w,x-w)=32i\pi,
\end{equation}
\begin{eqnarray}
\label{int3a5} &&\oint_C dx^{\mu}\int^x dy^{\nu}\int^y
dz^{\rho}\int^z dw^{\sigma}\int^w
dv^{\lambda}\epsilon_{\lambda\mu\alpha}\frac{(v-x)^{\alpha}}{\left|v-x
\right|^3}H_{\sigma\rho\nu}(z-w,y-w)\nonumber\\
&&~~~~~~~~~~~~~~~~~~~~~~~~~~~~~~~=32i\pi\frac{\pi^2}{6},
\end{eqnarray}
\begin{eqnarray}
\label{int3a6} &&\oint_C dx^{\mu}\int^x dy^{\nu}\int^y
dz^{\rho}\int^z dw^{\sigma}\int^w dv^{\lambda}\Bigg[\epsilon_{\lambda\sigma\alpha}\frac{(v-w)^{\alpha}}{\left|v-w\right|^3}H_{\rho\nu\mu}(y-z,x-z)\nonumber\\
&&~~~~+\epsilon_{\sigma\rho\alpha}\frac{(w-z)^{\alpha}}{\left|w-z\right|^3}H_{\lambda\nu\mu}(y-v,x-v)+\epsilon_{\rho\nu\alpha}\frac{(z-y)^{\alpha}}{\left|z-y
\right|^3}H_{\lambda\sigma\mu}(w-v,x-v)\nonumber\\
&&~~~+\epsilon_{\nu\mu\alpha}\frac{(y-x)^{\alpha}}{\left|y-x
\right|^3}H_{\lambda\sigma\rho}(w-v,z-v)\Bigg]=-32i\pi\frac{\pi^2}{6}.
\end{eqnarray}
More details of the calculation of
$\,(\ref{int3a1})$-$\,(\ref{int3a6})$ can be found in Appendix B.
By using the values of integrals in eqs.
$(\ref{int3a1})$-$\,(\ref{int3a6})$, we can calculate the VEV of
an unknotted Wilson loop operator for order $(\frac{2\pi}{k})^3$
in $\left\langle W_{\rho}(C)\right\rangle^{(3a)}$
\begin{equation}
\label{wo3a} \left\langle
W_{\rho}(\circlearrowleft)\right\rangle^{(3a)}=\frac{\:c_v^2\:\dim\:\rho\:c_{2}(\rho)}{k^3}\Bigg(\frac{2\pi^2}{3}\Bigg).
\end{equation}
The value of $\left\langle
W_{\rho}(\circlearrowleft)\right\rangle^{(3b)}$ is obtained by
using the  framing procedure as in eq. $\,(\ref{w2b})$
\begin{equation}
\label{wo3b} \left\langle
W_{\rho}(\circlearrowleft)\right\rangle^{(3b)}=0.
\end{equation}
From the equations $\,(\ref{dim})$, $\,(\ref{wcuo})$,
$\,(\ref{orde2auo})$, $\,(\ref{w2b})$, $\,(\ref{wo3a})$ and
$\,(\ref{wo3b})$, we can conclude that the calculation of VEV of
an unknotted Wilson loop operator up to order $(\frac{2\pi}{k})^3$
is given by
\begin{equation}
\label{wotot} \left\langle
W_{\rho}(\circlearrowleft)\right\rangle=\dim\:\rho\Bigg[1-\frac{1}{12}\Bigg(\frac{2\pi}{k}\Bigg)^2\:c_v\:c_2(\rho)+\frac{c_v^2\:c_2(\rho)}{3}\Bigg(\frac{2\pi^2}{k^3}\Bigg)+...\Bigg].
\end{equation}

We use the computation in the previous section for the gauge
groups $\mathop{\rm SU}(N)$ and $\mathop{\rm E}_6$ as examples.
For the gauge group $\mathop{\rm SU}(N)$, the values of dim$\rho$
and quadratic Casimir are
\begin{equation}
\label{dsun}
 \dim\:\rho=N,
\end{equation}
\begin{equation}
\label{c2sun} c_2(N)=Q(N)=\frac{N^2-1}{2N},
\end{equation}
\begin{equation}
\label{cvsun} c_v=Q(Adj)=N .
\end{equation}
Then, from non-perturbative case, we get
\begin{equation}
\label{esun}
E_0(N)=[N]_{\sqrt{q}}=N\Bigg[1-\frac{\pi^2}{k^2}\Bigg(\frac{N^2-1}{6}\Bigg)+\frac{2N\pi^2}{k^3}\Bigg(\frac{N^2-1}{6}\Bigg)+...\Bigg].
\end{equation}
If we calculate eq. $\,(\ref{wotot})$ by using the values in eqs.
$(\ref{dsun})$-$\,(\ref{cvsun})$, the VEV of an unknotted Wilson
loop operator for this gauge group, up to the same order, will
have the same values as in the equation $\,(\ref{esun})$:
\begin{equation}
\label{wosun} \left\langle
W_{N}(\circlearrowleft)\right\rangle=N\Bigg[1-\frac{\pi^2}{k^2}\Bigg(\frac{N^2-1}{6}\Bigg)+\frac{2N\pi^2}{k^3}\Bigg(\frac{N^2-1}{6}\Bigg)+...\Bigg].
\end{equation}
As in $\mathop{\rm SU}(N)$ case, the $\mathop{\rm E}_6$ group has
the following values :
\begin{equation}
\dim\:\rho=\dim\:27=27,
\end{equation}
\begin{equation}
c_2(27)=Q(27)=\frac{26}{3},
\end{equation}
\begin{equation}
c_v=Q(Adj)=Q(78)=12.
\end{equation}
From non-perturbative CSW theory, we get
\begin{equation}
\label{ee6}
E_0(27)=[3]_{q^2}[9]_{\sqrt{q}}=27-936\frac{\pi^2}{k^2}+22464\frac{\pi^2}{k^3}+....
\end{equation}
Then, as in eq. $\,(\ref{esun})$ and $\,(\ref{wosun})$, the
nonperturbative method in eq. $\,(\ref{ee6})$ will be identical to
the perturbative method in eq. $\,(\ref{woe6})$ up to order
$(\frac{2\pi}{k})^3$ of $\left\langle
W_{\rho}(\circlearrowleft)\right\rangle$ :
\begin{equation}
\label{woe6} \left\langle
W_{27}(\circlearrowleft)\right\rangle=27-936\frac{\pi^2}{k^2}+22464\frac{\pi^2}{k^3}+....
\end{equation}

\section{Conclusions and Discussions}

We have discussed the role of Wilson loop operators and extended
operators in the CSW theory. We have also discussed a two-particle
scattering  system, one of which we treat as a test particle
scattered off a source. In the calculation in
\cite{KoehlerMansouriVazWitten1991}, the second term of the
equation (\ref{scattering}), or the contributions from tetrahedron
operator, is missing. We evaluated this term for $\mathop{\rm
SU}(N)$ gauge group.

The calculation of the VEV of the Wilson loop operator in the CSW
theory has been discussed by Witten where he has showed that the
VEV of the Wilson loop operator in perturbation theory is the same
as the polynomial invariants of knot in three dimensions.

Looking at the $(\frac{1}{k})^0$ up to $(\frac{1}{k})^3$ terms of
the equation $\,(\ref{esun})$, $\,(\ref{wosun})$, $\,(\ref{ee6})$
and $\,(\ref{woe6})$, we summarize that the braiding formula is
identical up to the third order of the VEV of an unknotted Wilson
loop operator. For example, we have checked this result for the
gauge group $\mathop{\rm SU}(N)$ and $\mathop{\rm E}_6$. In fact,
our calculation showed that the symmetry and dynamical terms
factorize, so the contribution of the group factor can be computed
independently and hence the application of other gauge groups is
straightforward. Up to order $(\frac{1}{k})^2$, the VEV of the
Wilson loop operator has been computed in
\cite{GuadagniniMartelliniMintchev1989}. The problem arises in the
computation of the VEV of the Wilson loop operator for the
unknotted case of order $(\frac{1}{k})^3$. In this case, the use
of the framing procedure results in non-simple integral forms.

We hope that the result will help to illuminate more insights of
the equivalence between the braiding formula and the VEV of an
unknotted Wilson loop operator in perturbation theory and its
consistency with the equations $\,(\ref{esun})$ and
$\,(\ref{ee6})$ will strengthen this relation.

\section{Acknowledgements}
One of us (FPZ) would like to thank M. Hayashi for useful
discussions. AYW would like to thank BPPS, Dirjen Dikti, Republic
of Indonesia, for financial support. He also acknowledges all
members of Theoretical Physics Laboratory, Department of Physics
ITB, for warmest hospitality. This research is financially
supported by Riset Internasional ITB No. 054/K 01.07/PL/2008.




\newpage
\appendix
\section{Detailed Calculation of Ghost Contributions}
In the last section, we have calculate the contribution of ghost
fields to the VEV of the Wilson loop operator. In this appendix,
we provide more details of this calculation up to order $(1/k^2)$.

The first part of the eq.$\,(\ref{wrhoghost4})$ can be written as
\begin{equation}
\label{A2}\langle W_{\rho}(C)\rangle^{(2a)}_{ghost}=
\frac{\dim\:\rho}{4!}\Bigg[\int d^3
x\:f^{def}\Bigg(\partial^{\sigma}\frac{1}{\delta
H^d}\Bigg)\frac{\delta^3}{\delta J^{e\sigma}\delta{\bar
H}^f}\Bigg]^4\:Z_0\Bigg|_{J=H=\bar{H}=0}.
\end{equation}
Some terms of the solution of eq. $\,(\ref{A2})$ vanish since they
contain the determinant form with dependent columns or rows, i.e.
$\epsilon_{\alpha\beta\gamma}\epsilon_{\sigma\delta\lambda}.
p_1^{\alpha}p_1^{\beta}p_2^{\sigma}=0$. Therefore we can write eq.
$\,(\ref{A2})$ as
\begin{eqnarray}
\langle
W_{\rho}(C)\rangle^{(2a)}_{ghost}&=&-\dim\:\rho\Bigg(\frac{\pi}{2k\:c_2(\rho)}\Bigg)^2
f^{f_2e_1f_1}f^{f_4e_3f_2}f^{f_1e_3f_3}f^{f_3e_1f_4}
\epsilon_{\sigma_1\sigma_4\rho_1}\nonumber\\&&\times\epsilon_{\sigma_2\sigma_3\rho_2}
 \int d^3x\int\frac{d^3p_4}{(2\pi)^3}\frac{p_4^{\sigma_1}p_4^{\sigma_2}}{p_4^2}\int\frac{d^3p_5}{(2\pi)^3}\frac{p_5^{\rho_1}p_5^{\sigma_3}}{p_5^2(p_4-p_5)^2}\nonumber\\&&\times\int\frac{d^3p_6}{(2\pi)^3}\frac{p_6^{\rho_2}p_6^{\sigma_4}}{p_6^2(p_6-p_5)^2(p_6-p_5+p_4)^2}\nonumber\\&&-4.\frac{\dim\:\rho}{6}\Bigg(\frac{1}{64}\Bigg)\Bigg(\frac{\pi}{2k\:c_2(\rho)}\Bigg)^2
f^{f_2e_3f_1}f^{f_1e_4f_2}f^{f_4e_3f_3}f^{f_3e_4f_4}
\nonumber\\&&\times
\epsilon_{\sigma_1\sigma_3\rho_1}\epsilon_{\sigma_1\sigma_4\rho_2}\int
d^3x\int\frac{d^3p_4}{(2\pi)^3}\frac{p_4^{\rho_1}p_4^{\sigma_4}}{p_4^2}\int\frac{d^3p_5}{(2\pi)^3}\frac{p_5^{\rho_2}p_5^{\sigma_3}}{p_5^2(p_5-p_4)^3}\nonumber\\
&=&3.\dim\rho\Bigg(\frac{\pi}{16k\:c_2(\rho)}\Bigg)^2
f^{f_2e_1f_1}f^{f_4e_3f_2}f^{f_1e_3f_3}f^{f_3e_1f_4} \int
d^3x\epsilon_{\sigma_1\rho_2\rho_1}\nonumber\\&&\times\epsilon_{\sigma_2\sigma_3\rho_2}
 \int\frac{d^3p_4}{(2\pi)^3}\frac{p_4^{\sigma_1}p_4^{\sigma_2}}{p_4^5}\int\frac{d^3p_5}{(2\pi)^3}\Bigg[\frac{p_5^{\rho_1}p_5^{\sigma_3}}{(p_5+p_4)^2} -\frac{p_5^{\rho_1}p_5^{\sigma_3}}{p_5^2}\Bigg]\nonumber\\
 &&-\dim
\:\rho\Bigg(\frac{\pi}{2k\:c_2(\rho)}\Bigg)^2
f^{f_2e_1f_1}f^{f_4e_3f_2}f^{f_1e_3f_3}f^{f_3e_1f_4}
\epsilon_{\sigma_1\sigma_4\rho_1}\nonumber\\&&\times\epsilon_{\sigma_2\sigma_3\rho_2}
 \int
 d^3x\int\frac{d^3p_4}{(2\pi)^3}\frac{p_4^{\sigma_1}p_4^{\sigma_2}}{p_4^5}\int\frac{d^3p_5}{(2\pi)^3}\frac{p_5^{\rho_1}p_5^{\sigma_3}}{p_5^2(p_4-p_5)^2}\frac{1}{4\pi^2}\nonumber\\&&\times\Bigg[(p_4-p_5)^{\rho_2}(p_4-p_5)^{\sigma_4}B\Bigg(\frac{7}{2},1\Bigg)-p_5^{\rho_2}p_5^{\sigma_4}B\Bigg(\frac{7}{2},1\Bigg)\nonumber\\
&&-\frac{(p_4-p_5)^2}{4}\delta^{\rho_2\sigma_4}B\Bigg(\frac{5}{2},2\Bigg)+\frac{p_5^2}{4}\delta^{\rho_2\sigma_4}B\Bigg(\frac{5}{2},2\Bigg)\Bigg]\nonumber
\end{eqnarray}

\begin{eqnarray}
\label{A3}
&&+2\frac{\dim\:\rho}{3}\Bigg(\frac{\pi}{16k\:c_2(\rho)}\Bigg)^2\Bigg(\frac{1}{128}\Bigg)
f^{f_2e_3f_1}f^{f_1e_4f_2}f^{f_4e_3f_3}f^{f_3e_4f_4}
 \nonumber\\&&\times\epsilon_{\sigma_1\rho_2\rho_1}\epsilon_{\sigma_1\rho_2\sigma_4}
 \int
 d^3x\int\frac{d^3p_4}{(2\pi)^3}\frac{p_4^{\rho_1}p_4^{\sigma_4}}{p_4^2}\nonumber\\
 &=& 0.
 \end{eqnarray}
 The second part of eq. $\,(\ref{wrhoghost4})$ vanishes
 \begin{eqnarray}
\label{A4}\langle
W_{\rho}(C)\rangle^{(2b)}_{ghost}&=&\frac{1}{2!}\:c_2(\rho)\oint_{C}dx^{\tau}\int^x
dy^{\lambda}\frac{\delta^2}{\delta J^{i\tau}\delta
J^{i\lambda}}\nonumber\\&&~~~~\times\Bigg[\int d^3
x\:f^{def}\Bigg(\partial^{\sigma}\frac{1}{\delta
H^d}\Bigg)\frac{\delta^3}{\delta J^{e\sigma}\delta{\bar
H}^f}\Bigg]^2\:Z_0\Bigg|_{J=H=\bar{H}=0}\nonumber\\
&=&-\Bigg(\frac{\pi}{8k}\Bigg)^2\Bigg(\frac{c_v\:\dim\:\rho}{c_2(\rho)}\Bigg)\oint_c dx^{\tau}\int^x dy^{\lambda}\:\epsilon_{\sigma\lambda\alpha}\epsilon_{\sigma\tau\beta}\nonumber\\&&\times\int\frac{d^3p}{(2\pi)^3}\frac{p^{\alpha}p^{\beta}}{p^3}\cos[p.(x-y)]\nonumber\\
&=& 0,
\end{eqnarray}
because of the following relations
\begin{equation}
\label{A5}
\cos[p.(x-y)]=1-\frac{[p.(x-y)]^2}{2!}+\frac{[p.(x-y)]^4}{4!}.....,
\end{equation}
and
\begin{equation}
\label{A6}
\int\frac{d^3p}{(2\pi)^2}\frac{p^{\mu_1}p^{\mu_2}}{p^3}=\int\frac{d^3p}{(2\pi)^2}\frac{p^{\mu_1}p^{\mu_2}p^{\mu_3}}{p^3}=
... = 0.
\end{equation}

Finally, the third part of the eq.$\,(\ref{wrhoghost4})$ is
\begin{eqnarray}
\label{A7a} \langle
W_{\rho}(C)\rangle^{(2c)}_{ghost}&=&-\frac{\dim\:\rho}{3!}\Bigg[\int
d^3 x\:f^{def}\Bigg(\partial^{\sigma}\frac{1}{\delta
H^d}\Bigg)\frac{\delta^3}{\delta J^{e\sigma}\delta{\bar
H}^f}\Bigg]^3\nonumber\\
&&~\times\Bigg[\frac{k}{12\pi}\:c_2(\rho)\int d^3
x\:f^{abc}\epsilon^{\alpha\beta\gamma}\frac{\delta^3}{\delta
J^{a\alpha}\delta J^{b\beta}\delta
J^{c\gamma}}\Bigg]Z_0\Bigg|_{J=H=\bar{H}=0},
\end{eqnarray}
which, after eliminating the zero terms can be written as
\begin{eqnarray}
\langle
W_{\rho}(C)\rangle^{(2c)}_{ghost}&=&4\frac{\dim\:\rho\:k}{6\pi}\Bigg(\frac{\pi}{k}\Bigg)^3\Bigg(\frac{1}{2\:c_2(\rho)}\Bigg)^2
f^{f_3bf_1}f^{f_1cf_2}f^{f_2af_3}f^{abc}
\epsilon_{\sigma_2\rho_1\rho_2} \nonumber
\end{eqnarray}

\begin{eqnarray}
\label{A7} &&~~\times\epsilon_{\sigma_3\sigma_1\rho_2}\int
d^3x\Bigg[\int\frac{d^3p_1}{(2\pi)^3}\frac{p_1^{\sigma_1}p_1^{\sigma_2}p_1}{64}\int\frac{d^3p_2}{(2\pi)^3}\frac{p_2^{\sigma_3}p_2^{\rho_1}}{p_2^2(p_2-p_1)^6}\nonumber\\&&~~~~~~~~+\int\frac{d^3p_2}{(2\pi)^3}\frac{p_2^{\sigma_3}p_2^{\rho_1}p_2}{64}\int\frac{d^3p_1}{(2\pi)^3}\frac{p_1^{\sigma_1}p_1^{\sigma_2}}{p_1^2(p_1-p_2)^6}\Bigg]\nonumber\\
&&-4\frac{\dim\:\rho\:k}{18\pi}\Bigg(\frac{\pi}{k}\Bigg)^3
\Bigg(\frac{1}{2\:c_2(\rho)}\Bigg)^2
f^{f_3bf_1}f^{f_1cf_2}f^{f_2af_3}f^{abc}
\epsilon_{\sigma_2\rho_1\rho_2}
 \nonumber\\&&~~\times\epsilon_{\sigma_3\sigma_1\rho_3}\int d^3x\int\frac{d^3p_1}{(2\pi)^3}\frac{p_1^{\sigma_1}p_1^{\sigma_2}}{p_1^2}\int\frac{d^3p_2}{(2\pi)^3}\frac{p_2^{\sigma_3}p_2^{\rho_1}}{p_2^2(p_2-p_1)^2}\nonumber\\&&~~\times\frac{1}{4\pi^2}\frac{1}{(p_1-p_2)^3}\Bigg[p_1^{\rho_2}p_1^{\rho_3}B\Bigg(\frac{7}{2},1\bigg)-\frac{p_1^2}{4}\delta^{\rho_2\rho_3}B\Bigg(\frac{5}{2},2\Bigg)\nonumber\\&&~~~~~~-p_2^{\rho_2}p_2^{\rho_3}B\Bigg(\frac{7}{2},1\Bigg)+\frac{p_2^2}{4}\delta^{\rho_2\rho_3}B\Bigg(\frac{5}{2},2\Bigg)\Bigg]\nonumber\\
&&-\frac{\dim\:\rho\:k}{2\pi}\Bigg(\frac{\pi}{k}\Bigg)^3\frac{1}{64}\Bigg(\frac{1}{2\:c_2(\rho)}\Bigg)^2
f^{f_2bf_1}f^{f_3cf_2}f^{f_1af_3}f^{abc}
\nonumber\\&&~~\times\epsilon_{\sigma_1\rho_3\rho_1}\epsilon_{\sigma_2\sigma_3\rho_1}
 \int d^3x\int\frac{d^3p_1}{(2\pi)^3}\frac{p_1^{\sigma_1}p_1^{\sigma_3}}{p_1^5}\nonumber\\&&~~\times\Bigg[\int\frac{d^3p_2}{(2\pi)^3}\frac{p_2^{\sigma_2}p_2^{\rho_3}}{p_2^2}-\int\frac{d^3p_2}{(2\pi)^3}\frac{p_2^{\sigma_2}p_2^{\rho_3}}{(p_1+p_2)^2}\Bigg]\nonumber\\
&&+\frac{\dim\:\rho\:k}{6\pi}\Bigg(\frac{\pi}{k}\Bigg)^3\Bigg(\frac{1}{2\:c_2(\rho)}\Bigg)^2
f^{f_2bf_1}f^{f_3cf_2}f^{f_1af_3}f^{abc}
\epsilon_{\sigma_1\rho_3\rho_1}
 \nonumber\\&&~~\times\epsilon_{\sigma_2\sigma_3\rho_2}\int d^3x\int\frac{d^3p_2}{(2\pi)^3}\frac{p_2^{\sigma_2}p_2^{\rho_3}}{p_2^2}\int\frac{d^3p_1}{(2\pi)^3}\frac{p_1^{\sigma_1}p_1^{\sigma_3}}{p_1^5(p_1+p_2)^2}\frac{1}{4\pi^2}\nonumber\\&&~~\times\Bigg[(p_1+p_2)^{\rho_1}(p_1+p_2)^{\rho_2}B\Bigg(\frac{7}{2},1\Bigg)-p_2^{\rho_1}p_2^{\rho_2}B\Bigg(\frac{7}{2},1\Bigg)\nonumber\\&&~~~~~~-\frac{(p_1+p_2)^2}{4}\delta^{\rho_1\rho_2}B\Bigg(\frac{5}{2},2\Bigg)+\frac{p_2^2}{4}\delta^{\rho_1\rho_2}B\Bigg(\frac{5}{2},2\Bigg)\Bigg]\nonumber\\
&=& 0.
\end{eqnarray}

From the results of the above calculations, we can conclude that
the contributions of ghost fields to the VEV of the Wilson loop
operator up to order $(1/k^2)$ vanish.

\section{Detailed Calculation of The Integrals Using Framing}
In this paper, we use circle for the unknotted knot which
parametrization is given in eq.$\,(\ref{uo})$ and the vector field
orthogonal to $C$ is $n^{\mu}$ which is given in
eq.$\,(\ref{ns})$. Therefore we get
\begin{eqnarray}
\label{fram2}
&&\epsilon_{\mu\nu\sigma}\dot{x}^{\mu}(s)\left(\dot{x}^{\nu}(t)+\epsilon\dot{n}^{\nu}(t)\right)
(x(s)-x(t)-\epsilon n(t))^{\sigma}~~~~~~~~~~~~~~~~~~~\nonumber\\
&=&\det[\dot{x}(s)|\dot{x}(t)|x(s)-x(t)]+\epsilon\:
\det[\dot{x}(s)|\dot{n}(t)|x(s)-x(t)]\nonumber\\
& &-\epsilon\: \det[\dot{x}(s)|\dot{x}(t)|n(t)]-\epsilon^2\:
\det[\dot{x}(s)|\dot{n}(t)|n(t)]\nonumber\\
&=&\epsilon\: \det[\dot{x}(s)|\dot{n}(t)|x(s)-x(t)]-\epsilon\: \det[\dot{x}(s)|\dot{x}(t)|n(t)]\nonumber\\
&=&\epsilon\:e^{\pi i t}\Bigg[4\pi^4
i(s-t)^2-(2\pi)^2\Bigg(t-s-\frac{(t-s)^3}{3!}\Bigg)\Bigg].
\end{eqnarray}
By using the integral in the eq.$\,(\ref{fram})$, the imaginary
part of the integral is
\begin{eqnarray}
\label{intfr1} &&\int^t_0
du\:\epsilon_{\mu\nu\sigma}\dot{x}^{\mu}(s)\left(\dot{x}^{\nu}(u)+\epsilon\dot{n}^{\nu}(u)\right)
\frac{(x(s)-x(u)-\epsilon n(u))^{\sigma}}{\left|x(s)-x(u)-\epsilon
n(u)\right|^3}~~~~~~~~~~~~~~~~~~~~~~~~\nonumber\\
&=&\epsilon\:e^{\pi i s}\int^t_{s-\delta}du\frac{32\pi^4 i
(s-u)^2-32 \pi^2 (u-s)}{\left[16\pi^2
(u-s)^2+4\epsilon^2\right]^{3/2}}=-\pi\:e^{\pi i
s}\nonumber\\&=&-i\pi\:\sin\:\pi\:s,
\end{eqnarray}
\begin{eqnarray}
\label{intfr2} & &\int^t_0 du
\dot{x}^{\mu}(s)\dot{x}^{\nu}(t)\dot{x}^{\rho}(u)\:H_{\mu\nu\rho}(x(t)-x(s),x(u)-x(s))~~~~~~~~~~~~~~~~~\nonumber\\
&=&(2\pi)^4\int^t_0 du
\left[\sin\pi|s-t|+\sin\pi|s-u|+\sin\pi|t-u|\right]^{-1}\nonumber\\
&=&(2\pi)^3 \sec^2\Bigg[\frac{\pi
(s-t)}{2}\Bigg]\ln\Bigg(1-\cot\frac{\pi s}{2}\:\tan\frac{\pi
t}{2}\Bigg),
\end{eqnarray}
\begin{eqnarray}
\label{intfr3} &&\int^u_0 dg \int^g_0 dh\:
\dot{x}^{\sigma}(g)\dot{x}^{\nu}(t)\dot{x}^{\lambda}(h)\:H_{\lambda\sigma\nu}(x(g)-x(h),x(t)-x(h))\nonumber\\
&=&16\pi^2\int^u_{g=0}\ln\Bigg(1-\cot\frac{\pi t}{2}\tan\frac{\pi
g}{2}\Bigg)d\:\tan\Bigg[\frac{\pi(g-t)}{2}\Bigg]\nonumber\\
&=&16\pi^2\Bigg[\ln\Bigg(1-\cot\frac{\pi t}{2}\tan\frac{\pi
u}{2}\Bigg)\tan\Bigg(\frac{\pi
(u-t)}{2}\Bigg)\nonumber\\&&~~~~~~~~~-\cot\frac{\pi
t}{2}\ln\Bigg(1+\tan\frac{\pi t}{2}\tan\frac{\pi
u}{2}\Bigg)\Bigg].
\end{eqnarray}
Next, we will compute the integral in
eq.$\,(\ref{int3a1})$-$\,(\ref{int3a6})$ by framing the unknotted
knot.

The first, from the eq. $\,(\ref{int3a1})$, there is integral
upper limit $u \rightarrow t$ for integration variable $dg$ :

\begin{eqnarray}
\label{intAp1} &&\oint_C dx^{\mu}\int^x dy^{\nu}\int^y
dz^{\rho}\int^z dw^{\sigma}\int^w
dv^{\lambda}\:\epsilon_{\mu\rho\alpha}\frac{(x-z)^{\alpha}}{\left|x-z\right|^3}H_{\lambda\sigma\nu}(w-v,y-v)\nonumber\\
&=&\int_0^1 ds\int^s_0 dt\int^t_0 du\int^{u\rightarrow t}_0
dg\int^g_0 dh
\:\epsilon_{\mu\rho\alpha}\:\dot{x}^{\mu}(s)\left(\dot{x}^{\rho}(u)+\epsilon\dot{n}^{\rho}(u)\right)
\nonumber\\
&&\times\:\frac{(x(s)-x(u)-\epsilon
n(u))^{\alpha}}{\left|x(s)-x(u)-\epsilon
n(u)\right|^3}\dot{x}^{\nu}(t)\:\dot{x}^{\sigma}(g)\:\dot{x}^{\lambda}(h)\:H_{\lambda\sigma\nu}(x(g)-x(h),x(t)-x(h))\nonumber\\
&=&16\pi^3i\int^1_0 ds\: \sin\pi s\:\int^s_0 dt\: \cot\frac{\pi
t}{2}\:\ln\Bigg(1+\tan^2\frac{\pi t}{2}\Bigg)\nonumber\\
&=&32\pi i\Bigg(\frac{\pi^2}{6}-1\Bigg)
\end{eqnarray}
and from the eq. $\,(\ref{int3a2})$, there is integral limit $g
\rightarrow u$ for integration variable $dh$ :
\begin{eqnarray}
\label{intAp2} &&\oint_C dx^{\mu}\int^x dy^{\nu}\int^y
dz^{\rho}\int^z dw^{\sigma}\int^w
dv^{\lambda}\epsilon_{\mu\sigma\alpha}\frac{(x-w)^{\alpha}}{\left|x-w\right|^3}H_{\lambda\rho\nu}(z-v,y-v)\nonumber\\
&=&\int_0^1 ds\int^s_0 dt\int^t_0 du\int^u_0 dg\int^{g\rightarrow
u}_0
dh\:\epsilon_{\mu\sigma\alpha}\:\dot{x}^{\mu}(s)\left(\dot{x}^{\sigma}(g)+\epsilon\dot{n}^{\sigma}(g)\right)
\nonumber\\
&&\times\:\frac{(x(s)-x(g)-\epsilon
n(g))^{\alpha}}{\left|x(s)-x(g)-\epsilon
n(g)\right|^3}\dot{x}^{\nu}(t)\:\dot{x}^{\rho}(u)\:\dot{x}^{\lambda}(h)\:H_{\lambda\rho\nu}(x(u)-x(h),x(t)-x(h))\nonumber\\
&=&-8\pi^4i\int^1_0 ds\: \sin\pi s\:\int^s_0 dt\:
\int^t_0 du\:\sec^2\Bigg(\frac{\pi(u-t)}{2}\Bigg)\:\ln\Bigg(1-\cot\frac{\pi t}{2}\tan\frac{\pi u}{2}\Bigg)\nonumber\\
&=&32\pi i\Bigg(\frac{\pi^2}{6}-1\Bigg).
\end{eqnarray}
Then, from the eq.$\,(\ref{int3a3})$, we take the limit $u
\rightarrow t$ and $g \rightarrow u$ for integration variables
$dg$ and $dh$ respectively :
\begin{eqnarray}
&&\oint_C dx^{\mu}\int^x dy^{\nu}\int^y dz^{\rho}\int^z
dw^{\sigma}\int^w
dv^{\lambda}\Bigg[\epsilon_{\rho\lambda\alpha}\frac{(z-v)^{\alpha}}{\left|z-v
\right|^3}H_{\sigma\nu\mu}(y-w,x-w)\nonumber\\
&&~~~~~~~~~~~~~~~~~~~~~~+\epsilon_{\nu\sigma\alpha}\frac{(y-w)^{\alpha}}{\left|y-w
\right|^3}H_{\lambda\rho\mu}(z-v,x-v)\Bigg]\nonumber
\end{eqnarray}

\begin{eqnarray}
\label{intAp3} &=&\int_0^1 ds\int^s_0 dt\int^t_0
du\int^{u\rightarrow t}_0 dg\int^g_0 dh
\:\epsilon_{\rho\lambda\alpha}\:\dot{x}^{\rho}(u)\left(\dot{x}^{\lambda}(h)+\epsilon\dot{n}^{\lambda}(h)\right)
\nonumber\\
&&\times\:\frac{(x(u)-x(h)-\epsilon
n(h))^{\alpha}}{\left|x(u)-x(h)-\epsilon
n(h)\right|^3}\dot{x}^{\mu}(s)\:\dot{x}^{\nu}(t)\:\dot{x}^{\sigma}(g)\:H_{\sigma\nu\mu}(x(t)-x(g),x(s)-x(g))\nonumber\\
&+&\int_0^1 ds\int^s_0 dt\int^t_0 du\int^u_0 dg\int^{g\rightarrow
u}_0
dh\:\epsilon_{\nu\sigma\alpha}\:\dot{x}^{\nu}(t)\left(\dot{x}^{\sigma}(g)+\epsilon\dot{n}^{\sigma}(g)\right)
\nonumber\\
&&\times\:\frac{(x(t)-x(g)-\epsilon
n(g))^{\alpha}}{\left|x(t)-x(g)-\epsilon
n(g)\right|^3}\dot{x}^{\mu}(s)\:\dot{x}^{\rho}(u)\:\dot{x}^{\lambda}(h)\:H_{\lambda\rho\mu}(x(u)-x(h),x(s)-x(h))\nonumber\\
&=&16\pi^3i\int^1_0 ds\:\int^s_0 dt\:\Bigg[\frac{\tan^2(\pi
t/2)}{1+\tan\frac{\pi t}{2}\tan\frac{\pi s}{2}}+\nonumber\\&&~~~~~~~~~~~~~~~~~~~~+\sin\pi t\:\cot\Bigg(\frac{\pi s}{2}\Bigg)\:\ln\Bigg(1+\tan\frac{\pi t}{2}\tan\frac{\pi s}{2}\Bigg)\Bigg]\nonumber\\
&=&32\pi i,
\end{eqnarray}
and by using the limit $t \rightarrow s$ in the eq.
$\,(\ref{int3a4})$, we get
\begin{eqnarray}
\label{intAp4} &&\oint_C dx^{\mu}\int^x dy^{\nu}\int^y
dz^{\rho}\int^z dw^{\sigma}\int^w
dv^{\lambda}\epsilon_{\nu\lambda\alpha}\frac{(y-v)^{\alpha}}{\left|y-v\right|^3}H_{\sigma\rho\mu}(z-w,x-w)\nonumber\\
&=&\int_0^1 ds\int^s_0 dt\int^{t\rightarrow s}_0 du\int^u_0
dg\int^{g}_0
dh\:\epsilon_{\nu\lambda\alpha}\:\dot{x}^{\nu}(t)\left(\dot{x}^{\lambda}(h)+\epsilon\dot{n}^{\lambda}(h)\right)
\nonumber\\
&&\times\:\frac{(x(t)-x(h)-\epsilon
n(h))^{\alpha}}{\left|x(t)-x(h)-\epsilon
n(h)\right|^3}\dot{x}^{\mu}(s)\:\dot{x}^{\rho}(u)\:\dot{x}^{\sigma}(g)\:H_{\sigma\rho\mu}(x(u)-x(g),x(s)-x(g))\nonumber\\
&=&16\pi^3i\int^1_0 ds\:\int^s_0 dt\: \sin\pi t\:
\cot\frac{\pi s}{2}\:\ln\Bigg(1+\tan\frac{\pi s}{2}\tan\frac{\pi t}{2}\Bigg)\nonumber\\
&=&32\pi i.
\end{eqnarray}
Next, from the eq. $\,(\ref{int3a5})$, we take the limit $s
\rightarrow 1$ for integration variable $dt$ :
\begin{eqnarray}
\label{intAp5} &&\oint_C dx^{\mu}\int^x dy^{\nu}\int^y
dz^{\rho}\int^z dw^{\sigma}\int^w
dv^{\lambda}\epsilon_{\mu\lambda\alpha}\frac{(x-v)^{\alpha}}{\left|x-v
\right|^3}H_{\sigma\rho\nu}(z-w,y-w)\nonumber\\&=&\int_0^1
ds\int^{s\rightarrow 1}_0 dt\int^t_0 du\int^u_0 dg\int^{g}_0
dh\:\epsilon_{\mu\lambda\alpha}\:\dot{x}^{\mu}(s)\left(\dot{x}^{\lambda}(h)+\epsilon\dot{n}^{\lambda}(h)\right)
\nonumber\\
&&\times\:\frac{(x(s)-x(h)-\epsilon
n(h))^{\alpha}}{\left|x(s)-x(h)-\epsilon
n(h)\right|^3}\dot{x}^{\nu}(t)\:\dot{x}^{\rho}(u)\:\dot{x}^{\sigma}(g)\:H_{\sigma\rho\nu}(x(u)-x(g),x(t)-x(g))\nonumber\\
&=&-32i\pi\frac{\pi^2}{6},
\end{eqnarray}
and finally, the four integrals in equation $\,(\ref{int3a6})$ can
be evaluated as
\begin{eqnarray}
\label{intAp6} &&\oint_C dx^{\mu}\int^x dy^{\nu}\int^y
dz^{\rho}\int^z dw^{\sigma}\int^w dv^{\lambda}\nonumber\\
&&\times\Bigg[\epsilon_{\sigma\lambda\alpha}\frac{(w-v)^{\alpha}}{\left|w-v\right|^3}H_{\rho\nu\mu}(y-z,x-z)+\epsilon_{\rho\sigma\alpha}\frac{(z-w)^{\alpha}}{\left|z-w\right|^3}H_{\lambda\nu\mu}(y-v,x-v)\nonumber\\
&&+\epsilon_{\nu\rho\alpha}\frac{(y-z)^{\alpha}}{\left|y-z
\right|^3}H_{\lambda\sigma\mu}(w-v,x-v)+\epsilon_{\mu\nu\alpha}\frac{(x-y)^{\alpha}}{\left|x-y
\right|^3}H_{\lambda\sigma\rho}(w-v,z-v)\Bigg]\nonumber\\
&=&\int_0^1 ds\int^s_0 dt\int^t_0 du\int^u_0 dg\int^{g}_0
dh\:\dot{x}^{\mu}(s)\:\dot{x}^{\nu}(t)\:\dot{x}^{\rho}(u)\:\dot{x}^{\sigma}(g)\:\dot{x}^{\lambda}(h)\nonumber\\
&&\times\Bigg[\epsilon_{\sigma\lambda\alpha}\frac{(x(g)-x(h))^{\alpha}}{\left|x(g)-x(h)\right|^3}\:H_{\rho\nu\mu}(x(t)-x(u),x(s)-x(u))\nonumber\\&&+\epsilon_{\rho\sigma\alpha}\frac{(x(u)-x(g))^{\alpha}}{\left|x(u)-x(g)\right|^3}\:H_{\lambda\nu\mu}(x(t)-x(h),x(s)-x(h))\nonumber\\
&&+\epsilon_{\nu\rho\alpha}\frac{(x(t)-x(u))^{\alpha}}{\left|x(t)-x(u)\right|^3}\:H_{\lambda\sigma\mu}(x(g)-x(h),x(s)-x(h))\nonumber\\&&+\epsilon_{\mu\nu\alpha}\frac{(x(s)-x(t))^{\alpha}}{\left|x(s)-x(t)\right|^3}\:H_{\lambda\sigma\rho}(x(g)-x(h),x(u)-x(h))\Bigg]\nonumber\\
&=&-\pi\Bigg[\int_0^1 ds\int^s_0 dt\int^t_0
du\int^{u\rightarrow 1}_0 dg\:e^{\pi i g}\dot{x}^{\mu}(s)\:\dot{x}^{\nu}(t)\:\dot{x}^{\rho}(u)H_{\rho\nu\mu}(x(t)-x(u),x(s)-x(u))\nonumber\\
&&+\int_0^1 ds\int^s_0 dt\int^t_0 du\int^{g\rightarrow t}_0
dh\:e^{\pi i u}\dot{x}^{\mu}(s)\:\dot{x}^{\nu}(t)\:\dot{x}^{\lambda}(h)H_{\lambda\nu\mu}(x(t)-x(h),x(s)-x(h))\nonumber\\
&&+\int_0^1 ds\int^s_0 dt\int^{u\rightarrow s} _0 dg\int^{g}_0
dh\:e^{\pi i t}\dot{x}^{\mu}(s)\:\dot{x}^{\sigma}(g)\:\dot{x}^{\lambda}(h)H_{\lambda\sigma\mu}(x(g)-x(h),x(s)-x(h))\nonumber\\
&&+\int_0^1 ds\int^{t\rightarrow 1}_0 du\int^u_0 dg\int^{g}_0
dh\:e^{\pi i s}\:\dot{x}^{\rho}(u)\:\dot{x}^{\sigma}(g)\:\dot{x}^{\lambda}(h)H_{\lambda\sigma\rho}(x(g)-x(h),x(u)-x(h))\Bigg]\nonumber\\
&=&-\pi\Bigg[\int_0^1 ds\int^s_0 dt\int^t_0
du\int^{1}_0 dg\:e^{\pi i g}\dot{x}^{\mu}(s)\:\dot{x}^{\nu}(t)\:\dot{x}^{\rho}(u)H_{\rho\nu\mu}(x(t)-x(u),x(s)-x(u))\nonumber\\
&&+\int_0^1 ds\int^s_0 dt\int^t_0 du\int^{t}_0
dg\:e^{\pi i u}\dot{x}^{\mu}(s)\:\dot{x}^{\nu}(t)\:\dot{x}^{\sigma}(g)H_{\sigma\nu\mu}(x(t)-x(g),x(s)-x(g))\nonumber\\
&&+\int_0^1 ds\int^s_0 dt\int^{s}_0 du\int^{u}_0
dg\:e^{\pi i t}\dot{x}^{\mu}(s)\:\dot{x}^{\rho}(u)\:\dot{x}^{\sigma}(g)H_{\sigma\rho\mu}(x(u)-x(g),x(s)-x(g))\nonumber\\
&&+\int_0^1 ds\int^{1}_0 dt\int^t_0 du\int^{u}_0
dg\:e^{\pi i s}\:\dot{x}^{\nu}(t)\dot{x}^{\rho}(u)\:\dot{x}^{\sigma}(g)\:H_{\sigma\rho\nu}(x(u)-x(g),x(t)-x(g))\Bigg]\nonumber\\
&=&-\pi \frac{1}{4}\times 4\int_0^1 ds\int^s_0 dt\int^t_0
du\int^{1}_0 dg\:e^{\pi i g}\dot{x}^{\mu}(s)\:\dot{x}^{\nu}(t)\:\dot{x}^{\rho}(u)H_{\rho\nu\mu}(x(t)-x(u),x(s)-x(u))\nonumber\\
&=&-\pi i
(2\pi)^4\Bigg(\frac{1}{6\pi}\Bigg)\Bigg(\frac{2}{\pi}\Bigg)=-32\pi
i \Bigg(\frac{\pi^2}{6}\Bigg).
\end{eqnarray}

\bibliographystyle{plain}

\end{document}